\documentclass[a4paper]{article}
\usepackage{amsmath,amsfonts,amssymb,amsthm}

\newtheorem{theorem}{Theorem}[section]
\newtheorem{remark}[theorem]{Remark}

\newtheorem{lemma}[theorem]{Lemma}

\newcommand{\h}{{\cal H}}

\newcommand{\R}{{\mathbb R}}

\newcommand{\C}{{\mathbb C}}
\newcommand{\x}{{\bf x}}
\newcommand{\cst}{{ \mathcal{ C} }}
\newcommand{\A}{{\bf a}}
\newcommand{\e}{{\bf e}}
\newcommand{\tr}{{ \rm Tr }}

\newcommand{\y}{{\bf y}}
\newcommand{\z}{{\bf z}}
\newcommand{\fl}{{\rm fl}}
\newcommand{\Fl}{{\rm Fl}}
\begin{document}

\noindent  \centerline {\textbf{\Large Diamagnetic expansions for
    perfect quantum gases}\\}
\bigskip

\qquad \quad\qquad \qquad\qquad \today

\vspace{0.5cm}

\noindent \textbf{Philippe Briet
\footnote{PHYMAT-Universit\'e de Toulon et du Var, Centre de Physique
Th\'eorique-CNRS and FRUMAM,
Campus de Luminy, Case 907 13288 Marseille cedex 9, France; e-mail:briet@univ-tln.fr},
    Horia D. Cornean\footnote{Dept. of Math.,
    Aalborg
    University, Fredrik Bajers Vej 7G, 9220 Aalborg, Denmark; e-mail:
    cornean@math.aau.dk}, and Delphine Louis\footnote{Universit\'e de Toulon
et du Var,
PHYMAT-Centre de Physique Th\'eorique-CNRS and FRUMAM, Campus de Luminy, Case 907
                               13288 Marseille cedex 9, France;
                               e-mail:louis@cpt.univ-mrs.fr}}

\vspace{0.5cm}

\vskip3mm\noindent {\bf Abstract: }
In this work  we study the diamagnetic properties of a perfect quantum gas
in the presence of a  constant magnetic field of intensity $B$. We 
investigate the Gibbs semigroup 
 associated to the one particle operator at finite volume, and study
 its Taylor series with respect to the field parameter $\omega:= eB/c$
 in different topologies.  
This allows us to prove the existence of the thermodynamic limit for  
the pressure and for all its derivatives with respect to $\omega$ (the
so-called generalized susceptibilities).
\vspace{0.35cm}



\tableofcontents

\vskip3mm\noindent \section{Introduction and results.}
 This paper is  motivated  by the study of  the  diamagnetic properties of 
   a  perfect  quantum  gas  interacting with a constant magnetic field,   
${ \bf B}:= B {\bf e}_3, B>0$, ${\bf e}_3 :=(0,0,1)$. The  system obeys either  
the Bose or the Fermi statistics. Since we are only studying orbital
diamagnetic  effects, 
 we consider a gas of spinless  and
charged particles. 

We are mainly interesting in the bulk  response   i.e. the 
thermodynamic limit of the  pressure and its derivatives w.r.t.  cyclotron frequency
 $ \omega:= e B/c$. As in \cite{BCL}  we use the term {\it generalized
   susceptibilities} 
  to designate such quantities.

This question has been already addressed by several authors. Thus one finds results concerning the existence of
the  large volume limit of  the pressure for both Fermi and Bose gases \cite{AC, ABN2},  the  
magnetization for a Bose gas \cite{ C,MMP} and  the magnetic
susceptibility for a Fermi gas \cite{ABN2}.
 In \cite{BCL}, extensions of these results  to  the case of 
 generalized susceptibilities  were announced.

 This paper is the first in  a series of two devoted to the rigorous 
 proof  of the  results announced in \cite{BCL}. Here we consider the
 regime in which  the  inverse temperature   
$  \beta := 1/(kT) $  is positive and finite and  the  fugacity 
$z=e^{\beta\mu}$ belongs to the unit  complex disk.   Such  conditions were also used in \cite{AC,ABN2,MMP}. But
 here  we  also  allow   any positive   value of the cyclotron frequency
 $ \omega:= e/c B$. In a  forthcoming  paper we will  extend these results to  some larger 
$z-$complex domains  (in fact to ${\bf D}_\epsilon$ defined
below). One can  also find different aspects of this  problem in  \cite{MMP,CoRo,HeSj}.

 The main part of this work  is concerned with   a new  approach to the 
 magnetic perturbation theory for a semigroup generated by a magnetic
 Schr\"odinger operator. It extends the results given in  
 \cite{ABN1,C} and heavily relies on the use of the magnetic phase
factor.  This allows us to  have a good control
  on the magnetic  perturbation  w.r.t. the size of the volume
\cite{BC,CN} in which the gas is  confined.

Let us now describe our  results. Let  $\Lambda $  be an open, bounded and
connected subset of $\R^3$ containing  the origin of $\R^3$ and  with
smooth boundary $\partial \Lambda$.  Set
\begin{equation}\label{dome}
\Lambda_L:=\{\x\in \R^3,\; \x/L\in \Lambda\}; L>1.
\end{equation}
 Here we use
the  transverse gauge i.e. the magnetic potential is defined as  $
B\A:= (B/2)  {\bf e}_3 \wedge \x$.
The one particle Hamiltonian
\begin{equation}\label{hamunipa}
  H_L( \omega) = \frac{1}{2}(-i\nabla - \omega \A)^2,  
\end{equation}
 is first defined in the form sense on $H_0^1(\Lambda_L)$, and then
 one considers its Friedrichs extension \cite{RSII, RSIV}. Thus we work with Dirichlet
 boundary conditions (DBC). 

 It is well-known that  $H_{L}(\omega)$, $\omega \in \R$, 
generates  a  Gibbs semigroup  
\begin{equation}\label{gibbssemig}
 \{ W_L(\beta, \omega)= e^{-\beta
  H_{L}(\omega)}: \; 
    \beta \geq 0 \}
\end{equation}
i.e.   for all $\beta > 0$,  $W_L(\beta, \omega)\in B_1(L^2(\R^3)$, 
the  set of trace class operators
 on ${\cal H}_{L} $ \cite{AC,Z}.

Then for $ \beta >0$,
 $\omega \in  {\R}$, the grand canonical pressure of a quantum gas at finite
volume is defined as  (\cite{Hu, AC, ABN2}) 
 \begin{align} \label{pression}  P_L( \beta,\omega, z, \epsilon)= 
\frac{\epsilon }{\beta \vert \Lambda_L \vert }
 \cdot \tr \left\{\ln( 1+ \epsilon z  W_L(\beta, \omega) \right\},
\end{align}
where  $ \epsilon =-1$ ($\epsilon =1$) for
Bose (Fermi) statistics. Since $\omega/2=\inf \sigma(H_\infty( \omega))$,
then the pressure is  an analytic   function w.r.t. $z $ on the complex domain  ${\bf D}_\epsilon$ with
$$ {\bf D}_{+1}:= {\C} \setminus (-\infty, -e^{\beta
\omega/2}], \quad {\bf D}_{-1}:={\C} \setminus [e^{\beta \omega/2},\infty).$$
Let $ n \geq 1$ and  define the susceptibility of order $n$  at finite volume by:
\begin{equation}  \label{defchi}
\chi_L^{(n)}( \beta,\omega, z, \epsilon):= \frac{\partial^n P_L} {\partial\omega^n }( \beta,\omega, z,
\epsilon);  
\end{equation}
If $n=0$ we set $\chi_L^{(0)}( \beta,\omega, z, \epsilon):=P_L( \beta,\omega, z, \epsilon)$.

Our first result describes the properties of the above defined
quantities at finite volume,  and it  is given  by the following
theorem:

\begin{theorem}  \label{finitevolume} Let $\beta> 0$. Then the map $\R\ni \omega \to W_L(\beta, \omega)\in
B_1(L^2(\R^3))$ is real analytic and admits an entire extension. 
For each open and bounded set $\mathcal{K}$ which obeys $\overline{\mathcal{K}}\subset {\bf D}_\epsilon$, $\epsilon= -1,+1$, 
there exists an open neighborhood ${\cal N}$ of the real
axis such that  the pressure at finite volume  $P_L(\beta, \omega,z, \epsilon)$ is 
analytic w.r.t.  $ (\omega,z) $ on $ {\cal N} \times \mathcal{K}
$. 
 Let $ \omega \in \R$, and $ \vert z \vert < 1$. Then for
 $n \geq 0$ we have (see \eqref{gibbssemig}):
 \begin{eqnarray} \label{chiL}
\chi_L^{(n)}(\beta, \omega,z,\epsilon)=   \frac{\epsilon}{\beta \vert
  \Lambda_L \vert}\sum_{k\geq 1}
\frac{(-\epsilon z)^k }{k}  \quad { \tr} \left \{ \frac{\partial^n
    W_L(k\beta, \omega)}{ \partial \omega^n} \right \}.
\end{eqnarray}
\end{theorem}

 We now  discuss the limit $L= \infty$. First we 
 define  the candidates, $ \chi_\infty^{(n)}$, for these limits.
 Recall that the one particle operator   $H_\infty(\omega)=
 \frac{1}{2}(-i\nabla-\omega \A)^2$ on $L^2({\R}^3)$, $\omega\in\R$,  is
positive and  essentially selfadjoint on $C^\infty_0
({\R}^3)$. Denote by $W_\infty(\beta, \omega), \beta \geq 0 $ the
semigroup generated by $H_\infty(\omega)$.  Then $W_\infty(\beta, \omega)$ has an 
explicit integral kernel satisfying (see Section 3):
\begin{equation} \label{Ginfinidiag}
G_\infty(\x,\x; \beta,\omega)=  \frac{ 1 }{(2 \pi \beta)^{3/2}}\frac{ \omega \beta/2}{ \sinh( \omega \beta/2
)},\quad \forall \x\in\R^3.
\end{equation}
Note that the right hand side is independent of $\x$. Let $ \beta >0,
\omega \geq 0$ and $\vert z \vert<1$. In view of \eqref{chiL}, define:
\begin{equation} \label{Pinfini}
P_\infty(\beta, \omega,z,\epsilon):= \frac{\epsilon}{\beta}\sum_{k\geq
  1}\frac{(-\epsilon z)^k }{k}  G_\infty({\bf 0},{\bf 0}; k\beta,\omega),
\end{equation}
which is well defined because of the estimate $\sinh (t) \geq t $ if $t \geq 0$.
 Then   by the results of \cite{AC, ABN2}, we know that  
$$\lim_{L \to \infty}P_L(\beta, \omega,z,\epsilon) = P_\infty(\beta, \omega,z,\epsilon).$$

It is quite  natural to choose $\chi_\infty^{(n)}:= {\frac{\partial^n P_\infty}{\partial\omega^n }}$ provided that this last quantity exists.
Note that it is not very easy to see this just from \eqref{Pinfini}  and \eqref{Ginfinidiag}. 

 We  then prove the following 
\begin{theorem}  \label{infinitevolume}
Let $\beta >0$, $\omega \geq 0$  and $ \vert z\vert <1$. Fix $n\geq1$
and define 
\begin{equation} \label{chiinfini1}
\chi_\infty^{(n)}(\beta, \omega,z,\epsilon):= {\frac{\partial^n P_\infty}
{\partial\omega^n }}( \beta,\omega, z,
\epsilon).
\end{equation}
Then we have the equality : 
\begin{equation} \label{chiinfini}
\chi_\infty^{(n)}(\beta, \omega,z,\epsilon)= \frac{\epsilon}{\beta}\sum_{k\geq0}
\frac{(-\epsilon z)^k }{k}   \frac{\partial^n G_\infty}{ \partial \omega^n} ({\bf 0},{\bf 0}; k\beta,\omega) .
\end{equation}
Moreover, 
\begin{equation} \label{limite}
 \lim_{L \to \infty }  \chi_L^{(n)} ( \beta, \omega,z,\epsilon )=  \chi_\infty^{(n)}( \beta, \omega,z,
\epsilon)
\end{equation}
 uniformly on $ [\beta_0, \beta_1] \times  [\omega_0,\omega_1]$, $0<\beta_0 < \beta_1<\infty$ 
and $ 0< \omega_0 <
\omega_1 <
\infty$.
\end{theorem}

\subsection{Relation with the de Haas-van
  Alphen (dHvA) effect }

Our results can be easily extended to the case of more general Bloch
electrons, that is when one has a background smooth and periodic electric
potential $V$. More precisely, let us assume that $V\in C^\infty(\R^3)$,
$V\geq 0$, and if $\Gamma$ is a periodic
lattice in $\R^3$ then $V(\cdot)=V(\cdot +\gamma)$ for all $\gamma\in
\Gamma$. Denote by $\Omega$ the elementary cell of $\Gamma$. In this
case, the grandcanonical pressure at the thermodynamic limit will be given
by (we work with fermions thus $\epsilon=1$)
\begin{equation} \label{Pinfini22}
P_\infty(\beta, \omega,z)= \frac{1}{\beta}\sum_{k\geq
  1}\frac{(- z)^k }{k} \frac{1}{|\Omega|}\int_\Omega G_\infty(\x,\x; k\beta,\omega) d\x,
\end{equation}
where $G_\infty(\x,\x'; k\beta,\omega)$ is the smooth integral kernel
of the semigroup generated by $\frac{1}{2}(-i\nabla+\omega
\A)^2+V$. This formula only holds for $|z|<1$, but it can be
analytically continued to $\C\setminus (-\infty,-1]$, see \cite{AC} or
\cite{HeSj}. 

Now one can start looking at the behavior of $P_\infty(\beta,
\omega,z)$ as function of $\omega$, in particular around the point
$\omega_0=0$. Working in canonical conditions, that is when $z$ is a
function of $\beta$, $\omega$ and the fixed particle density $\rho$,
then one is interesting in the object 
$$p_\infty(\beta,\omega,\rho):=P_\infty(\beta, \omega,z(\beta,\omega,\rho)).$$ 
A thorough analysis of the $\omega$ behavior near $0$, involving
derivatives with respect to $\omega$ of the above quantity, has been already
given by Helffer and Sj\"ostrand in \cite{HeSj}. 

Alternatively, one can start from the finite volume quantities, and
define a  $z_L(\beta,\omega,\rho)$ as the unique solution of the
equation $\rho_L:=\beta z\partial_z P_L(\beta,\omega,z_L)=\rho$ 
and $p_L(\beta,\omega,\rho):=P_\infty(\beta, \omega,z_L(\beta,\omega,\rho))$. Is it still true
that at large volumes we have for example that 
$$\partial_\omega^n p_L(\beta,\omega,\rho)\sim \partial_\omega^n
p_\infty(\beta,\omega,\rho),\quad n\geq 1\;?$$
The main achievement of our paper is that at least
for small densities (which fix $|z|<1$) the answer is yes. In a companion paper we will
prove that this is true for all $z\in \C\setminus (-\infty,-1]$.

\bigskip

 We end the introduction by giving the plan of this paper. In Section 2 we discuss the analyticity of the Gibbs
semigroup with respect to $ \omega$  in the trace class sense. The
trace norm estimates we obtain depend on the size of the domain, due to the
linear growth of the magnetic potential. Using magnetic perturbation
theory we manage to regularize the trace expansions and to extend these results to the infinite
volume case in Sections 3 and 4. Finally we prove the existence
of thermodynamic limits in Section 5.

\section{ Analyticity of Gibbs semigroups}\label{section2}
\setcounter{equation}{0}
\subsection {$B_1$-Analyticity}
Let  $\Lambda_L$, $L\geq 1$  be domains of $\R^3$ as defined in
\eqref{dome}. In the following we will denote respectively  by $\Vert T\Vert_1$, 
 $\Vert T\Vert_2$ and  $ \Vert T\Vert $,  the trace norm in $B_1(L^2(\Lambda_L))$, the
 Hilbert-Schmidt norm in $B_2(L^2(\Lambda_L))$ and the operator norm
 in $B(L^2(\Lambda_L))$ of $T$. 

 In this section  we study   the   $ \omega$ expansion of $W_L(\beta,
 \omega)$. This question has been already considered \cite{HP,ABN2,Z}
 in connection with the ${B}_1$ analyticity of $W_L(\beta, \omega)$. Combining their result 
  with  our analysis below, this gives 
   the following.  Define the operators:
\begin{align} 
\hat{R}_{1,L}(\beta,\omega) &:=  \A\cdot(i\nabla_\x+\omega
\A)W_L(\beta,\omega),\\ 
\hat{R}_{2,L}(\beta,\omega) &:= \label{op R}
\frac{1}{2}\;\A^2\,W_L(\beta,\omega).
\end{align} 
 Both operators $\hat{R}_{1,L}(\beta,\omega)$ and
 $\hat{R}_{2,L}(\beta,\omega)$ belong to $B(\Lambda_L)$ and we
 have the following estimate on their norm:
\begin{lemma} \label{normeR}
For all  $ \beta >0$,  $\omega \geq 0$ and $L>1$, there exists a
positive constant $C$ such that:
\begin {equation} \label{normesop}
\|\hat R_{1,L}\|\leq \, \frac{C\, L}{\sqrt{\beta}}\quad{\rm and }\quad 
\|\hat R_{2,L}\|\leq \, C\, L^2.
\end{equation}
 \end{lemma}

\proof Let  $\varphi \in L^2(\Lambda_L)$. Since $ W_L(\beta,\omega)
L^2(\Lambda_L) \subset {\rm Dom}(H_{L}(\omega))$ \cite{K}, we have
after a standard argument (note that the absolute value of the components of $\A$ are
bounded from above by ${\rm diam}(\Lambda_1)\cdot L$): 
\begin{align}\nonumber
 \Vert  \A\cdot(i\nabla_\x+\omega \A)W_L(\beta,\omega) \varphi\Vert^2
 & \leq  C\, L^2 
  \left \langle H_L(\omega) 
 W_L(\beta,\omega) \varphi,W_L(\beta,\omega) \varphi\right \rangle \nonumber \\ 
&\leq \frac{C\cdot L^2}{\beta} \Vert   \varphi\Vert^2,
\end{align} 
 where the last estimate is given by the spectral theorem. The second bound of \eqref{normesop} is obvious. \qed 

\vspace{0.5cm}


{\remark  Due to the diamagnetic inequality  (see \eqref{diam} below),
 we have for all $\beta >0$ and  $\omega \in \R$:
\begin{equation}\label {WL1}
\Vert W_L(\beta,\omega)  \Vert_{1}  ={\rm Tr}\,
  \left (W_L(\beta,\omega)\right ) \leq  \frac{L^3}{(2\pi \beta)^{3/2}}.
\end{equation}
 Then both operators $\hat{R}_{1,L}, \hat{R}_{2,L}$  are trace class,
 since we can factorize the operator $\hat{R}_{1,L}(\beta,\omega)= \hat{R}_{1,L}(\beta/2,\omega)  W_L(\beta/2,\omega)$.}

\vspace{0.5cm}

For $n\geq 1$ define: 
\begin{equation} \label{Dn}
{\cal D}_n(\beta):= \{ 0<\tau_n  < \tau_{n-1}< ... \tau_1< \beta \}\subset \R^n.
\end{equation} 
Let $(i_1,...,i_n)\in\{1,2\}^{n}$. Lemma \ref{normeR} allows us  to define the following  family 
of bounded operators:
\begin{align}\label{defInL}
&
{\hat I}_{n,L}(i_1,...,i_n)(\beta,\omega):=
\int_{ {\cal D}_n(\beta)}\,W_L(\beta-\tau_1,\omega)\hat{R}_{i_1,L}(\tau_1-\tau_2,\omega)\\
&  \cdot
\hat{R}_{i_2,L}(\tau_2-\tau_3,\omega)...
\hat{R}_{i_{n-1},L}(\tau_{n-1}-\tau_n,\omega)\hat{R}_{i_n,L}(\tau_n,\omega)d\tau,\nonumber
\end{align}
where  $d\tau$ is the $n$-dimensional Lebesgue measure. These
operators are in fact trace class, and we will estimate 
their trace norm later. Let $n \geq 1$, $(i_1,...,i_n)\in\{1,2\}^{n}$ and  $\chi_k^{n}$ be the characteristic function,
\begin{eqnarray}\label{fctcar}
\chi_k^{n}(i_1,...,i_k):=\left\{\begin{array}{ccc} 1 &&{\rm if}\,\,
    i_1+...+i_k=n \\ 0 && {\rm  otherwise .}\end{array}\right.
\end{eqnarray}
Then we have
\begin{theorem}\label{thm1}
Fix $\beta >0$. Then the operator-valued function $\R\ni \omega\mapsto W_L(\beta,\omega)\in
{B}_1$ admits an entire extension to the whole complex plane.  Fix $\omega_0 \geq 0$. For all
$\omega\in\C$ we have
\begin{align}\label{deriveeWL}
W_L(\beta,\omega) &=
  \sum_{n=0}^{\infty}\frac{(\omega-\omega_0)^n}{n!}\, \, \frac{\partial^n
  W_L}{\partial \omega^n}(\beta,\omega_0), \\\label{akasa1}
\frac{1}{n!}\frac{\partial^n
  W_L}{\partial \omega^n}(\beta,\omega_0) &=
 \sum_{k=1}^{n}(-1)^k\sum_{i_j\in\{1,2\}}
\chi_k^n(i_1,...,i_k) {\hat I}_{k,L}(i_1,...,i_k)(\beta,\omega_0).
\end{align}
Moreover, there exists a positive constant $C$ independent of $n\geq 1$,
$\beta >0$ and 
$L$ such that:
\begin{equation} \label{norm1der}
 \left \Vert \frac{1}{n!}\frac{\partial^n
  W_L}{\partial \omega^n}(\beta,\omega_0)\right \Vert_{1} \leq  C^n \;\frac{(1+\beta)^{n}}{\beta^{3/2}}
L^{n+3} \frac{1}{[(n-1)/4]!}.
\end{equation}
For all $\omega \in \C$, $ \{W_L(\beta,\omega), \beta > 0\} $ is a Gibbs semigroup
 with its generator given  by the closed operator $H_{L}(\omega)$.
\end{theorem}
\begin{remark} This   theorem implies  that the trace of the semigroup $W_L$ 
 is an entire function of  $\omega $ and by \eqref{deriveeWL}:
\begin{equation}
\tr \left(  W_L(\beta,\omega)\right)=  \sum_{n=0}^{\infty}\frac{(\omega-\omega_0)^n}{n!}\,  
\tr \left(\frac{\partial^n
  W_L}{\partial \omega^n}(\beta,\omega_0) \right).
\end{equation}
\medskip
\end{remark}

\noindent {\it Proof of Theorem \ref{thm1}.}  
We will use here some results from \cite{HP} and  \cite{ABN2}, which
we briefly recall.  Let $\omega_0 \geq 0$.
For $ \omega \in \C $ set $\delta \omega := \omega-\omega_0  $.  Then   the operator
\begin{equation} \label{Rchap}
 H_{L}(\omega)- H_{L}(\omega_0)=   (\delta \omega)\,  \A\cdot(i\nabla_\x+\omega_0
\A) +  \frac{(\delta \omega)^2}{2}\;\A^2
\end{equation} is relatively bounded
to $H_{L}(\beta,\omega_0)$ with relative bound zero. Note that for $\beta>0$
 from  \eqref{Rchap} we have  in the  operator sense on ${\cal H}_L$ 
\begin{align}\nonumber
\hat{R}_{L}(\beta,\omega) &:= \left( H_{L}(\omega)- H_{L}(\omega_0)
\right)W_L(\beta,\omega_0)
 \nonumber \\  
&=\delta\omega\;\hat{R}_{1,L}(\beta,\omega_0) + (\delta\omega)^2 \hat{R}_{2,L}(\beta,\omega_0).\label{def-RL}
\end{align} 
 For every compact subset $ K\subset\C$, and due to the estimates
 \eqref{normesop}, this operator satisfies 
$$\int_0^1 d\tau\,\sup_{\omega\in K}\| \hat{R}_L(\tau,\omega)\|<\infty.$$
   Let $0< \beta_1 \leq \beta_0< \infty$. Then the series 
\begin{equation}\label{serie-ABN}
W_L(\beta,\omega,\omega_0)=\sum_{n=0}^\infty (-1)^n W_L^{(n)}(\beta,\omega,\omega_0)
\end{equation}
where  $W_L^{(0)}(\beta,\omega,\omega_0)= W_L(\beta,\omega_0)$ and for  $n\geq 1$
\begin{equation} \label {coefABN}
W_L^{(n)}(\beta,\omega,\omega_0) = \int_0^{\beta}d\tau\,W_L(\beta-\tau,\omega_0) \left(
 H_{L}(\omega)- H_{L}(\omega_0)\right)W_L^{(n-1)}(\tau,\omega_0)
\end{equation}
is  uniformly $B_1$- convergent   on $K \times [\beta_1,
\beta_0]$. 
This result was obtained in \cite{ABN2}. Since
$W_L(\beta,\omega,\omega_0)$ is the uniform limit of a sequence of
entire $B_1$-valued functions, it follows via the Cauchy integral
formula that $W_L(\beta,\omega,\omega_0)$ is also  $B_1$-entire in
$\omega$. Moreover, for real $\omega$ it coincides with the operator
   $e^{-\beta H_L(\omega)}$, $\beta >0$. 

What we do here is to identify its $n$th order derivative with respect
 to $\omega$.  From
\eqref{coefABN} and \eqref{def-RL} a simple induction argument  yields
 the following finite rearranging:
\begin{align}
&\!\!\!\!\!\!\!\!\sum_{n=0}^N(-1)^{n} W_L^{(n)}(\beta,\omega, \omega_0) 
=W_L(\beta,\omega_0)+\sum_{n=1}^{N}
(\delta\omega)^n\sum_{k=1}^{n}(-1)^k\sum_{i_j\in\{1,2\}}\chi_k^n(i_1,...,i_k)
\nonumber \\ &
\!\!\!\!\!\!\!\! \cdot \hat I_{k,L}(i_1,...,i_k)(\beta,\omega_0)+
{\cal R}_{N+1,L}(\beta,\omega, \omega_0),\label{coefABN1}
\end{align}
 where
\begin{align}\label{restABN1}
& {\cal R}_{N+1,L}(\beta,\omega, \omega_0) \\ 
&=\sum_{n=N+1}^{2N}
(\delta\omega)^n\sum_{k=1}^{N}(-1)^k\sum_{i_j\in\{1,2\}}\chi_k^n(i_1,...,i_k) 
\hat I_{k,L}(i_1,...,i_k)(\beta,\omega_0).\nonumber 
\end{align}
Now differentiation with respect to $\omega$ commutes with the limit
$N\to\infty$, again due to the uniform convergence and the Cauchy
integral formula. Hence  \eqref{deriveeWL} is proved, since the $n$th
order derivative of $\sum_{j=0}^N(-1)^{j} W_L^{(j)}(\beta,\omega,
\omega_0)$ at $\omega=\omega_0$ equals the right hand side of
\eqref{akasa1} if $N\geq n$.
 
 In the second part of the proof, we use the methods of \cite{ABN2} in
 order to estimate the $B_1$-norm of the operators
 $I_{k,L}(i_1,...,i_k)$ as claimed in \eqref{norm1der}.  We first have:
\begin{align} \label{restABN20}
\Vert \hat I_{k,L}(i_1,...,i_k)(\beta,\omega_0)\Vert_{1} \leq  &\\ 
\int_{ {\cal D}_k(\beta)}d\tau &\, \Vert W_L(\beta-\tau_1,\omega_0)
\hat{R}_{i_1,L}(\tau_1-\tau_2,\omega_0)
...\hat{R}_{i_k,L}(\tau_k,\omega_0) \Vert_{1}. \nonumber 
\end{align}
 Recall  that the Ginibre-Gruber inequality read as  \cite{ABN2},
\begin{equation}\label{Ginibre-Gruber}
\Vert \prod^k_{l=0} A_l T(t_l)\Vert_{1} \leq \left( \prod^k_{l=0} \Vert  A_l  \Vert \right) 
\tr \; T(t_0+t_1 +...+t_k)
\end{equation} 
  where $\{A_l, 0 \leq l\leq k\}$ are  bounded operators
 and $T(t),t>0$ is a Gibbs semigroup. Then 
taking $A_0:=W_L( \frac{\beta-\tau_1}{2},\omega_0)$, 
$A_l:=\hat{R}_{i_l,L}( \frac{\tau_l-\tau_{l+1}}{2},\omega_0) $
 if $l\geq 1$ (we put $\tau_{k+1}\equiv 0$) and $T(t)=W_L( \frac{t}{2},\omega_0)$.  On $D_k(\beta)$, we have the 
estimate $\Vert  A_0  \Vert \leq 1 \leq \,
 \sqrt {\frac{ 1+\beta}{\beta-\tau_1}}$ and   by the Lemma \ref{normeR} for
$l\geq 1$
\begin {equation} \label{normR}
\Vert  A_l  \Vert = \Vert \hat{R}_{i_l,L} (\frac{\tau_l-\tau_{l+1}}{2},
 \omega_0)\| \leq {\rm const}\cdot \,
 L^{i_l}\frac{ \sqrt{1+\beta} }{ \sqrt{\tau_l-\tau_{l+1}}}.
\end{equation}
 Let
$f_k:{\cal D}_k(\beta) \to \R$ defined as
\begin{equation} \label{f}
f_k(\tau):=\frac {1}{ \sqrt{( \beta-\tau_1)( \tau_1-\tau_2)...(\tau_{k-1}-\tau_k)\tau_k}}, \;\;
\end{equation}
and it  satisfies                                        
\begin{equation} \label{f1}
\int_{{\cal D}_k(\beta)}f_k(\tau) d \tau =
 \frac{\beta^{\frac{k-1}{2}}\pi^{\frac{k+1}{2}}}{\Gamma\left(\frac{k+1}{2}\right)}.
\end{equation}
  Let $i_1+...+i_k=n$. Then  from \eqref{WL1}, \eqref{Ginibre-Gruber},
 \eqref{normR} and  \eqref{f1}, we obtain the existence of a numerical
 constant $C$, such that for every $\beta >0$:
\begin{eqnarray} \label{restABN2}
\Vert \hat I_{k,L}(i_1,...,i_k)(\beta,\omega_0)\Vert_{1} \leq  
 \frac {L^{n+3}C^k(1+\beta)^{k}}{\beta^{3/2}\Gamma\left(\frac{k+1}{2}\right)} .
\end{eqnarray}
 Thus we have the estimate (see \eqref{akasa1}):
\begin {equation} \label{restABN3}
 \frac{1}{n!}\Vert (\partial_\omega^n W_L)(\beta,\omega_0) \Vert_{1} \leq C^n L^{n+3}\frac{(1+\beta)^{n}}{\beta^{3/2}}
\sum_{k=1}^{n}\sum_{i_j\in\{1,2\}}
\frac{\chi_k^n(i_1,...,i_k) }{\Gamma\left(\frac{k+1}{2}\right)}.
\end{equation}
But a lot of terms in the above sum are zero, since
$\chi_k^n(i_1,...,i_k)=0$ if $ k < [\frac{n+1}{2}]$. Since $\Gamma$ is
increasing, we can give a rough estimate of the form:
\begin{equation}\label{gammaesti}
\sum_{k=1}^{n}\sum_{i_j\in\{1,2\}}
\frac{\chi_k^n(i_1,...,i_k) }{\Gamma\left(\frac{k+1}{2}\right)}\leq
n\; 2^n\frac{1 }{\Gamma\left(\frac{[(n+1)/2]+1}{2}\right)}\leq n\; 2^n\frac{1 }{[(n-1)/4]!}.
\end{equation}

\qed


\subsection{  Proof of Theorem \ref{finitevolume}.}

The analyticity properties of the pressure are now easy to prove once we have the $B_1$-analyticity
 of the Gibbs semigroup. See \cite{ABN1} for details. 

Now let $ \beta>0, \omega \geq 0$ and $\vert z\vert <1$.  Since $
\Vert z W_L(\beta,\omega) \Vert <1$, the logarithm in the pressure at
finite volume can be expanded and then:
\begin{align} \label{PL}
P_L(\beta, \omega,z,\epsilon)= \epsilon/(\beta \vert \Lambda_L
\vert)\sum_{k\geq 1}
(-\epsilon z)^k /k  \; \tr\;  W_L(k\beta,\omega) .
\end{align}
Starting from Definition \eqref{defchi}, and using Theorem \ref{thm1} we obtain: 
\begin{eqnarray} \label{PL2}
\chi_L^{(n)}(\beta, \omega,z,\epsilon)=\epsilon/(\beta \vert \Lambda_L \vert)\sum_{k\geq0}
(-\epsilon z)^k /k  \quad { \tr} \left ( \frac{\partial^n
    W_L(k\beta,\omega)}{ \partial \omega^n} \right ).
\end{eqnarray}
Note that \eqref{norm1der} insures that the growth in $k$ which comes
from the trace of the $n$th derivative of $W_L(k\beta,\omega)$ is not
faster than some polynomial, but since $|z|<1$ the series in $k$ is
convergent. This finishes the proof of the theorem. \qed
\subsection{ Analyticity of the semigroup's integral kernel}

In the rest of this paper we will only consider $\Lambda_L= (-L/2,L/2)^3, L\geq 1 $.  For  $ \omega \in \R$,
 $H_{L}(\omega)$ is essentially selfadjoint on
$$ \{ \phi \in  C^1(\bar \Lambda_L) \cup C^2(\Lambda_L),
 \phi\vert_{\partial\Lambda_L}=0, \Delta \phi \in L^2(\Lambda_L)\}.$$

 Let $G_L(\x,\x';\beta,\omega)$ be the integral kernel of
$W_L(\beta, \omega)$ (see e.g. \cite {AC}). Standard elliptic
estimates for the eigenfunctions of $H_L(\omega)$, together with the
fact that $e^{-\beta H_L(\omega)}$ is trace class imply that
$G_L(\x,\x';\beta,\omega)$ is smooth in $ (\x, \x') \in
\Lambda_L\times \Lambda_L$. Moreover,  $G_L(\x,\x';\beta,\omega)=0$ if either $\x$ or
$\x'$ are on the boundary.  

To prove the next theorem, we need  the following result from
\cite{C}, concerning the $C^1$ regularity up to the boundary of the
integral kernel. Let $\beta>0$ and let 
$ G_\infty(\x,\x', \beta):= G_\infty(\x,\x', \beta,\omega=0)$ be the
heat kernel on the whole space, i.e.
\begin{equation} \label{heatk}
 G_\infty(\x,\x'; \beta) = \frac{1}{(2\pi \beta)^{3/2}} e^{- \frac{\vert \x-\x' \vert^2}{2\beta}}.
\end{equation}
 Recall that the diamagnetic estimate reads as \cite{AC}:
\begin{equation} \label{diam}
\vert G_L(\x,\x';\beta,\omega)\vert \leq   G_\infty(\x,\x'; \beta), \quad 
 (\x,\x') \in \Lambda_L\times \Lambda_L, \quad \omega \geq 0.
\end{equation}
 Then we have
\begin{lemma}\label{lm1} Let $\beta>0 $ and $\omega \geq 0$. Then on
  $\Lambda_L \times \Lambda_L$ we have: 
\begin{equation}\label{estdG}
 \vert(i\nabla_\x+\omega
\A(\x))G_L(\x,\x';\beta,\omega)\vert \leq  \frac{ \cst}{\sqrt \beta} \, G_\infty(\x,\x',
8\beta).
\end{equation}
where $ \cst=\cst(\beta,\omega)=   c \cdot (1+\beta)^{5}(1+\omega)^3 $ and $  c>1$
 is a numerical  constant.
\end{lemma}

This estimate allows us to define the integral kernels of the
operators defined in \eqref{op R}; more precisely, for $(\x,\x') \in
\Lambda_L\times \Lambda_L$ we have:
\begin{eqnarray}\nonumber
\hat{R}_{1,L}(\x,\x';\beta,\omega) &:=&  \A(\x)\cdot(i\nabla_\x+\omega
\A(\x))G_L(\x,\x';\beta,\omega),\\  \label{def-R1L-R2L}
\hat{R}_{2,L}(\x,\x';\beta,\omega) &:=&
\frac{1}{2}\;\A^2(\x)\,G_L(\x,\x';\beta,\omega).
\end{eqnarray}

Consider the operator $W_L(\beta,\omega)$ for complex $\omega$,
defined by a $B_1$-convergent complex power series in Theorem \ref{thm1}. We
will now prove that it has an integral kernel analytic in $\omega$:
\begin{theorem}\label{thm2}
Let $\beta>0$ and fix $\omega_0\geq 0$.

\noindent {\rm (i)}. The operator
$(\partial_\omega^nW_L)(\beta,\omega_0)$ defined in \eqref{akasa1} has
 an integral kernel denoted by
 $(\partial_\omega^nW_L)(\x,\x';\beta,\omega_0)$, 
 which is jointly continuous on $(\x,\x')\in\overline{\Lambda}_L\times
 {\overline\Lambda}_L$, and obeys the estimate 
 \begin {equation} \label{estf}
 \frac{1}{n!}\left \vert \frac{\partial^n W_L}{\partial
\omega^n}( \x,\x'; \beta,\omega_0)\right \vert  \leq c^n \frac
{(1+\omega_0)^{3n}(1+\beta)^{6n}L^n} {\beta^{3/2}\; [\frac {n-1}{4}]!} 
\quad n\geq 1
\end{equation}
 for some numerical constant $c \geq 1$.

\noindent {\rm (ii)}. For $\omega\in \C$, the operator $W_L(\beta,\omega)$ has an
integral kernel $G_L(\x,\x';\beta,\omega)$ given by:
\begin{equation} \label{expker}
  G_L(\x,\x';\beta,\omega) =  \sum_{n=0}^{\infty}\frac{(\omega-\omega_0)^n}{n!}\, \, 
\left(\frac{\partial^n
  W_L}{\partial \omega^n}\right )(\x,\x';\beta,\omega_0),
\end{equation}
where the above series is uniformly convergent on $\overline{\Lambda}_L\times
 {\overline\Lambda}_L$. Thus $G_L$ is jointly continuous  on $\bar \Lambda_L\times\bar \Lambda_L $
  and  is an entire function of $\omega$.
\end{theorem}


\medskip

 {\it Proof of  the Theorem \ref{thm2}}. Lemma \ref{lm1} obviously
 implies for $\beta  >0$ and  $\omega  \geq 0$ the estimate: 
\begin{equation}\label{lm10}
\vert \hat{R}_{1,L}(\x,\x'; \beta,\omega)\vert \leq  L\frac{  \cst} {\sqrt \beta} 
 G_\infty(\x,\x'; 8\beta).
\end{equation}
We also have
\begin{equation}\label{lm11}
 \vert \hat{R}_{2,L}(\x,\x'; \beta,\omega)\vert \leq  \frac{ L^2}{4}  G_\infty(\x,\x'; \beta).
\end{equation}
In the following,  we will often use the uniform estimate with respect
to the index  $i=1,2$: 
\begin{equation}\label{lm12}
 \vert \hat{R}_{i,L}(\x,\x'; t,\omega)\vert \leq 
  L^i \cst_1 \sqrt{\frac{(1+\beta)}{t}}G_\infty(\x,\x'; 
8t) \quad 0 <t \leq \beta, 
\end{equation}
where   $\cst_1:= \cst_1(\beta,\omega)= 2\sqrt2 \cst(\beta,\omega) $.

 Let us start by proving $(i)$. Fix $\beta>0, \omega_0 \geq 0, L \geq
 1$, and consider 
 the operator $\hat I_{k,L}(i_1,...,i_k)(\beta,\omega_0) $,
 $k\geq 1$ 
defined in \eqref{defInL}. It admits a continuous integral kernel, $\hat I_{k,L}(\x,\x';\beta,\omega_0):=
\hat I_{k,L}(i_1,...,i_k)(\x,\x';\beta,\omega_0)$
 on $ \Lambda_L \times \Lambda_L$
given by

\begin{align}\label{defInL1}
&\hat I_{k,L}(\x,\x';\beta,\omega_0) \\
&=\int_{D_k(\beta)}d\tau\int_{\Lambda_L^k}d\y
G_L(\x,\y_1;\beta-\tau_1,\omega_0)\hat{R}_{i_1,L}(\y_1,\y_2;\tau_1-\tau_2,\omega_0) \nonumber\\
& \cdot\dots \cdot 
\hat{R}_{i_{k-1},L}(\y_{k-1},\y_n;\tau_{k-1}-\tau_k,\omega_0) 
\hat{R}_{i_k,L}(\y_k,\x';\tau_k,\omega_0)\nonumber
\end{align}
where $d\y$ denotes the Lebesgue measure on $\R^{3k}$ and  $D_k(\beta)$
 is defined in \eqref{Dn}. Let $i_1 +i_2 +...+i_k=n$.
Then by using the Lemma \ref{lm1}, the estimate 
$$ \vert G_L(\x,\y_1;\beta-\tau_1,\omega_0) \vert \leq 
8^{3/2}\sqrt{\frac{ \beta} {\beta-\tau_1}}G_
\infty(\x,\y_1;8(\beta-\tau_1)),\qquad 0< \tau_1< \beta,$$ and \eqref{lm12}, the
following  estimate holds on
$\Lambda_L \times \Lambda_L$:
\begin{eqnarray} \nonumber &{}&
\vert  \hat I_{k,L}(\x,\x';\beta,\omega_0)\vert \leq 
8^{3/2}(1+ \beta) ^{(k+1)/2}\cst_1^k L^n\int_{D_k(\beta)}f_k(\tau)
d\tau\int_{\Lambda_L^k}d\y
 \\ \label{estInL1O}
&{}& 
G_ \infty(\x,\y_1;8(\beta-\tau_1))...G_ \infty(\y_k,\x';8\tau_k),  
\end{eqnarray}
where  the function $f_k$ is defined in \eqref{f}. Notice that by using  the semigroup property
 \begin{align}  \nonumber
 \int_{\Lambda_L^k}d\y G_ \infty(\x,\y_1;t_1))...G_ \infty(\y_k,\x';t_k)  &\leq 
\int_{\R^k}d\y G_ \infty(\x,\y_1;t_1))...G_ \infty(\y_k,\x';t_k)\\
\label{estnoy}
&=G_ \infty(\x,\x';t_1+...+t_k)
\end{align}
Therefore  from \eqref{estInL1O} we get 
\begin{equation}\label{estInL1}
\vert  \hat I_{k,L}(\x,\x';\beta,\omega_0)\vert \leq  
8^{3/2} (1+ \beta) ^{(k+1)/2}  \cst_1^k \, L^n  G_ \infty(\x,\x';8\beta)  
\int_{D_k(\beta)}f_k(\tau)d\tau.
\end{equation}
 Then Theorem \ref{thm1} together with
 \eqref{f1}  and \eqref{estInL1} show that the operator 
$\frac{\partial^n  W_L}{\partial \omega^n}(\beta,\omega_0), n\geq 1$  given by \eqref{deriveeWL}, 
admits a continuous integral kernel  satisfying
$$ \vert \frac{\partial^n
  W_L}{\partial \omega^n}( \x,\x';\beta,\omega_0) \vert \leq 
(8^3\pi)^{1/2} \cst_2^n  n! L^n \, G_\infty(\x,\x';8\beta)
  \sum_{k=1}^{n}\sum_{i_j\in\{1,2\}}
\frac{ \chi_k^n(i_1,...,i_k)   }{\Gamma\left(\frac{k+1}{2}\right)}.
$$
for a new constant $ \cst_2= \cst_2 (\beta,\omega_0) 
:= \pi^{1/2}(1+ \beta) \cst_1(\beta,\omega_0)$.
Then  by mimicking  the proof of \eqref{gammaesti} we get from 
the last inequality
\begin{equation} \label{estdwn}
\frac{1}{n!}\left \vert \frac{\partial^n
  W_L}{\partial \omega^n}( \x,\x';\beta,\omega_0)
 \right \vert \leq \frac{c^n (1+\beta)^{6n}(1+\omega_0)^{3n} \, L^n} { [(n-1)/4]!}
G_\infty(\x,\x';8\beta),
\end {equation}
where $c$ is a numerical constant. 
Since  $G_\infty(\x,\x';8\beta) \leq  (16\pi\beta)^{-3/2}$,\eqref{estdwn}  implies the estimate \eqref{estf} 
and  proves $(i)$. Then $(ii)$ follows easily from the previous
estimate since $1/[(n-1)/4]!$ has a super-exponential decay in $n$. \qed 

\section{Regularized expansion}
\setcounter{equation}{0}
The bounds obtained  in  the previous section are not convenient for 
the proof of the existence   of the thermodynamic limit of the  magnetic susceptibilities.
  In particular the  bound on $\frac{\partial^n
  W_L}{\partial \omega^n}( \x,\x',\beta,\omega_0) $ given by  \eqref{estdwn}  
 is of  order  $L^n$. Then  this gives a bound on its  trace 
of order $L^{3+n}$, while  in view of \eqref{chiL} we need a bound
which goes like $L^{3}$.

In this section, we  give an improvement of these estimates. In order to  do  that  we need to   introduce the magnetic
 phase $\phi$ and the magnetic flux $fl$ defined as (here $\x,\y,\z\in\Lambda_L$ and   ${\bf e}=(0,0,1)$):
\begin{align}\label{phaseflux}
 \phi(\x,\y)&:=\frac{1}{2}\,{\bf e}\cdot (\y\wedge
\x)=-\phi(\y,\x), \\\label{fluxyi}
\fl(\x,\y,\z)&:=\phi(\x,\y)+\phi(\y,\z)+
\phi(\z,\x)=\frac{1}{2}\,{\bf e}\cdot \{(\x-\y)\wedge (\z-\y)\}.
\end{align}
Note that $fl$ is really the magnetic flux through the triangle
defined by the three vectors, and we have: 
\begin{equation} \label{estfl}
|\fl(\x,\y,\z)|\leq\,|\x-\y|\, |\y-\z|.
\end{equation}
For  $n\geq 1$ and 
$ \x=\y_0$, $\y_1,..., \y_n$ some arbitrary vectors in $\Lambda_L$,  define
\begin{align} \label{defFL}
{\Fl}_n(\x,\y_1,...,\y_n)&:= \phi(\y_n,\x)+\sum_{k=0}^{n-1} \phi(\y_k,\y_{k+1})
\\
&= \sum_{k=1}^{n-1}\fl (\x,\y_k,\y_{k+1}),\;
 \rm{if}\quad  n\geq 2  \nonumber 
 \\
& \quad  \rm{and }  \quad {\Fl}_1(\x,\y_1)=0.\nonumber  
\end{align}
Notice that  due to \eqref{estfl}, we have
\begin{equation} \label {estFL}
\vert {\Fl}_n(\x,\y_1,...,\y_n)\vert \leq \sum_{k=1}^{n-1}
 \sum_{l=1}^{k}\vert\y_{l-1}-\y_{l}\vert \vert\y_{k}-\y_{k+1} \vert.
\end{equation}

Let  $\omega \geq 0$. Consider now the bounded operators given by their integral
kernels on $ \Lambda_L \times \Lambda_L$,
\begin{eqnarray}\label{def_R1L-R2L-RL}
R_{1,L}(\x,\x';\beta,\omega)& := & \A(\x-\x')\cdot
\left(i\nabla_\x+\omega
  \A\left(\x\right)\right)G_L(\x,\x';\beta,\omega).\nonumber \\
 R_{2,L}(\x,\x';\beta,\omega)& := & \frac{1}{2}\,
\A^2(\x-\x')G_L(\x,\x';\beta,\omega).
\end{eqnarray} 
Then by the Lemma \ref{lm1}, a straightforward estimate   yields to
\begin{equation}\label{est_R1L-R2L-RL0}
 \vert R_{1,L}(\x,\x';\beta,\omega) \vert \leq  \frac{ \cst \vert \x-\x'\vert }{2\sqrt \beta}
G_\infty(\x,\x',8\beta) \leq 4\cst_1 G_\infty(\x,\x',16\beta)
\end{equation} 
 for all  $(\x,\x') \in \Lambda_L \times \Lambda_L$. Similarly by \eqref{diam}we have
\begin{equation}\label{est_R1L-R2L-RL1}
 \vert R_{2,L}(\x,\x';\beta,\omega) \vert \leq  \frac{ \vert \x-\x'\vert^2 }{8}
G_\infty(\x,\x',\beta) \leq   \frac{\beta}{\sqrt2  }  G_\infty(\x,\x',2\beta).
\end{equation} 
In the sequel for $i=1,2$ we will use the  estimate on $\Lambda_L \times \Lambda_L$
\begin{equation}\label{est_R1L-R2L-RL}
 \vert R_{i,L}(\x,\x';\beta,\omega) \vert \leq  \cst_3 G_\infty(\x,\x',16\beta)
\end{equation} 
where $\cst_3= \cst_3(\beta,\omega):= 16  \cst_1(\beta,\omega) $ and $ \cst_1$
 is given in \eqref{lm12}.

Notice that \eqref{est_R1L-R2L-RL} provides an uniform bound w.r.t. $L$ and $\beta$ near $\beta =0$
 on the operator kernels. This  in contrast with   the  bound on the norm operator of $\hat R_{i,L};
i=1,2$ (see section 2.2, 
\eqref{lm10} and \eqref{lm11}).  Using the Schur-Holmgren estimate for
the operator norm of an integral operator,  \eqref{est_R1L-R2L-RL} 
 eventually implies  
\begin{equation} \label{nrR1R2}
\Vert R_{i,L}\Vert \leq \cst_3, \quad i=1,2.
\end{equation}
Let $\x\in\Lambda_L$. For $k\geq 1, m\geq 0$, $ \omega \geq 0$ and
$\beta>0$,   define the continuous function
\begin{align}\nonumber
&
W_{k,L}^m(\x;\beta,\omega):=\sum_{j=1}^k(-1)^j\sum_{(i_1,...,i_j)\in\{1,2\}^{j}}\chi_{j}^{k}(i_1,...,i_j)
\int_{D_j(\beta)}d\tau\int_{\Lambda^j}d\y\\ \nonumber
& 
\frac{\left(i\left(\Fl_{j}(\x,\y_1,...,\y_{j})\right)\right)^m}{m!}\,\,G_L(\x,\y_1;\beta-\tau_1,\omega)
R_{i_1,L}(\y_1,\y_2;\tau_1-\tau_2,\omega)\\
\label{def-WL-n-k}
& 
\,...\,R_{i_{j-1},L}
(\y_{j-1},\y_j;\tau_{j-1}-\tau_j,\omega)R_{i_j,L}(\y_j,\x;\tau_j,\omega),
\end{align}
 where   in the case of
$m=0$ we set $0^0\equiv 1$.
\medskip

The  main result  of this section  gives a new expression for the diagonal of kernel's
$n$th derivative with respect to $\omega$ at finite volume.
\begin{theorem}\label{thm3}
Let $\beta>0$ and $\omega\geq 0$. Then for all $\x\in\Lambda_L$, and for all $n\geq 1$,
one has
\begin{align}\label{deriveenn}
\frac{1}{n!}\,\frac{\partial^nG_L}{\partial\omega^n}(\x,\x,\beta,\omega)=\sum_{k=1}^n
W_{k,L}^{n-k}(\x;\beta,\omega),
\end{align}
and moreover, uniformly in $L>1$:
\begin{equation}\label{upperbo}
|W_{k,L}^m(\x;\beta,\omega)|\leq c(m,k) \frac{(1+
  \beta)^{7(m+k)+3}}{\beta^{3/2}}(1+ \omega)^{3(m+k)+2},
\end{equation}
where $c(m,k)=
 c^{m+k}\; \frac {m^{m}}{m!}\sum_{j=1}^{k}  
\frac{ j^{2m}}{j!}$ and $c$ is again a numerical factor. 
\end{theorem}

\noindent {\it Proof.} We first need to introduce some new notation. Fix $\omega_0 \geq 0$.
Let   $\omega \in \C$, $\delta\omega = \omega-\omega_0$ and   $\widetilde R_{i,L}(\beta,\omega, \omega_0),
 i=1,2$, $\widetilde W_L(\beta,\omega, \omega_0)$ be 
the operators on ${\cal H}_{L}$ defined  via  their respective integral kernel given on $ \Lambda_L \times \Lambda_L$, by
\begin{align}\nonumber
\widetilde R_{i,L}(\x,\x';\beta,\omega,\omega_0)&= e^{i\delta\omega\phi(\x,\x')}
R_{i,L}(\x,\x';\beta,\omega_0),\,\,\,i=1,2,  \\\label{uaidtilda}
\widetilde W_L(\x,\x';\beta,\omega,\omega_0)&= e^{i\delta\omega\phi(\x,\x')}G_L(\x,\x';\beta,\omega_0)
\end{align}
 where  $\phi$ is  defined in \eqref{phaseflux}. We also set
$$ \widetilde R _{L}(\x,\x';\beta,\omega,\omega_0):=\delta\omega  \widetilde 
R_{1,L}(\x,\x';\beta,\omega,\omega_0) + (\delta\omega)^2  \widetilde 
R_{2,L}(\x,\x';\beta,\omega,\omega_0). $$
Except a phase factor the kernel of $\widetilde W_L $ and  $ \widetilde 
R_{i,L}, i=1,2$ is   the   same as the one of $ W_L $,  $ 
R_{i,L}$, $i=1,2$  respectively. Then they satisfy \eqref{diam},  \eqref{est_R1L-R2L-RL}  respectively. 
Hence  by  the same arguments  as above,  they are bounded  operators 
   and 
\begin{equation} \label{normt}
\Vert  \widetilde W_{L}\Vert \leq 1, \, \Vert  \widetilde R_{i,L}\Vert \leq \cst_3
\end{equation}
(see \eqref{nrR1R2}). Notice also that since  $ \Vert \widetilde  W_{L}\Vert_{{\cal I}_2} 
= \Vert  W_{L}\Vert_{{\cal I}_2}\leq \frac {L^{3/2}}{(2\pi \beta)^{3/4}}$
  then by \eqref{est_R1L-R2L-RL}
  $ R_{i,L},i=1,2$ as well as  $ \widetilde 
R_{i,L},i=1,2$  are in the  Hilbert-Schmidt class and for
 $\beta >0$, $\omega_0 \geq 0$ and $ \omega \in \C$,
 \begin{equation} \label{nHS}
\Vert  R_{i,L}(\beta, \omega)\Vert_{{\cal I}_2}, \,  \Vert 
\widetilde  R_{i,L}\Vert_{{\cal I}_2}(\beta, \omega, \omega_0)
\leq \cst_3\Vert  W_{L}(16\beta, \omega)\Vert_{{\cal I}_2} 
\leq  {\cst_3}\frac {L^{3/2}}{(2\pi \beta)^{3/4}}.
\end{equation}
 where $\cst_3=\cst_3( \beta, \omega_0)$ was first introduced in
 \eqref{est_R1L-R2L-RL}. 
We now define the following family of  bounded  operators on ${\cal H}_L$.
Let $k \geq 1$, $\{i_1,...,i_k\}\in\{1,2\}^k$, $\beta>0$, $\omega_0 \geq 0$. For all $\omega \in \C$ set
\begin{eqnarray}\nonumber
&&  {\widetilde I_{k,L}}(i_1,...,i_k)(\beta,\omega,\omega_0)
:=\int_{D_k(\beta)}d\tau\,\widetilde W_L(\beta-\tau_1,\omega, \omega_0)
\widetilde R_{i_1,L}(\tau_1-\tau_2,\omega,\omega_0) \\ \label{InL2-def}
&& ...\widetilde R_{i_{k-1},L}(\tau_{k-1}-\tau_k,\omega,\omega_0)
\widetilde R_{i_k,L}(\tau_k,\omega,\omega_0),
\end{eqnarray}
and for 
$n \geq 1$  
\begin{eqnarray}\label{def-WLN}
W_{n,L}(\beta,\omega,\omega_0):=\sum_{k=1}^n(-1)^k \sum_{i_j\in\{1,2\}}\chi_{k}^{n}
(i_1,...,i_k){\widetilde I}_{k,L}(i_1,...,i_k)
(\beta,\omega,\omega_0)
\end{eqnarray}
where  $\chi_{k}^{n}$ is  defined in \eqref{fctcar}.
\begin{lemma} \label{lm3}
Let $N \geq 1$, $\beta>0$, $\omega_0 \geq 0$. For all $\omega \in \C$ set 
 $\delta\omega= \omega-\omega_0$. Then as bounded operators we can write:
\begin{eqnarray} \label{pseudoWL0}
W_L(\beta,\omega)= \widetilde W_L(\beta,\omega,\omega_0) +\sum_{n=1}^{N} (\delta\omega)^{n}\,
W_{n,L}(\beta,\omega,\omega_0) + \\ \nonumber   { \cal  \widetilde R}^{(1)}_{N+1,L}
(\beta, \omega,\omega_0) + \,{ \cal  \widetilde
R}^{(2)}_{N+1,L}(\beta,\omega,\omega_0),
\end{eqnarray}
where 
${ \cal  \widetilde R}^{(1)}_{N+1,L}(\beta, \omega,\omega_0)$ and $ 
{  \cal \widetilde R}^{(2)}_{N+1,L}(\beta, \omega,\omega_0) $
 are  the
following  bounded operators on $\h_L$:
\begin{align} \nonumber
& {\cal \widetilde R}^{(1)}_{N+1,L}(\beta,\omega,\omega_0) :=
(-1)^{N+1}\sum_{n=N+1}^{2N+2}(\delta\omega)^n\sum_{i_j\in\{1,2\}}\chi_{N+1}^n(i_1,...,i_n)\\ 
&\cdot \int_{D_{N+1}(\beta)}d\tau
\,W_L(\beta-\tau_1,\omega){\cal\widetilde
R}_{i_1,L}(\tau_1-\tau_2,\omega,\omega_0) 
  ... {\cal \widetilde R}_{i_{N+1},L}(\tau_{N+1},\omega,\omega_0)
\label{restreg1} \end{align} 
 where   $D_{N+1}(\beta)$ is  given in \eqref{Dn} and 
\begin{align}  \label{restreg2}
&{\cal \widetilde R}^{(2)}_{N+1,L}(\beta,\omega,\omega_0)\\
&=\sum_{n=N+1}^{2N}(\delta\omega)^n\sum_{k=1}^{N}(-1)^k\sum_{i_j\in\{1,2\}}\chi_k^n(i_1,...,i_k)  
 \cdot {\widetilde I_{k,L}}(i_1,...,i_k)(\beta,\omega,\omega_0).\nonumber
\end{align}
\end{lemma}
\noindent {\it Proof of the lemma.}  In this proof we fix $\omega_0 \geq 0$ and omit 
 everywhere the $\omega_0$ dependence.  We first note that 
$\widetilde W_L(\beta,\omega)$ is strongly differentiable with respect to 
$\beta > 0$  (see \cite{C}) and  satisfies
$$ \frac{\partial \widetilde W_L(\beta,\omega)}{\partial \beta} +
 H_L(\omega)\widetilde W_L(\beta,\omega) =
{ \widetilde R}_{L}(\beta,\omega).$$
By using Proposition 3 from \cite{C},  we can write the following Dyson-type
integral equation:
\begin{eqnarray}\label{perturb-semi-groupe-bis}
W_L(\beta,\omega)=\widetilde W_L(\beta,\omega)-\int_{0}^{\beta}d\tau\,W_L(\beta-\tau,\omega)
\widetilde  R_L(\tau,\omega).
\end{eqnarray}
The above integral is a Riemann integral and converges in the operator
norm sense. By iterating \eqref{perturb-semi-groupe-bis} we obtain:
\begin{align}\nonumber
&W_L(\beta,\omega)=\widetilde W_L(\beta,\omega)+\sum_{n=1}^{N}(-1)^{n}\int_0^{\beta}
d\tau_1\int_0^{\beta-\tau_1}d\tau_2...\int_0^{\beta-\tau_1-...-\tau_{n-1}}d\tau_n\nonumber \\
& \cdot\widetilde W_L (\beta-\tau_1-...-\tau_n,\omega)\widetilde  R_L(\tau_n,\omega)
...\widetilde  R_L(\tau_1,\omega)+ {\cal \widetilde R}^{(1)}_{N+1,L}(\beta,\omega)   
\end{align}
 where  
\begin{eqnarray}
{\cal \widetilde R}^{(1)}_{N+1,L}(\beta,\omega)=   (-1)^{N+1}\int_0^{\beta}
d\tau_1\int_0^{\beta-\tau_1}\!d\tau_2\,...\int_0^{\beta-\tau_1-...-\tau_N}
d\tau_{N+1}\nonumber \\\label{refintermed}
\cdot W_L(\beta-\tau_1-...-\tau_{N+1},\omega) \cdot{ \widetilde R}_L(\tau_{N+1},\omega)..
.{{\widetilde R}_L}(\tau_1,\omega). 
\end{eqnarray}
Then a straightforward change of variables  in the integrals of the r.h.s. of  the last two formulas,
yields to
\begin{eqnarray*}
&& W_L(\beta,\omega)=\widetilde W_L(\beta,\omega)+
\sum_{n=1}^{N}(-1)^{n}\int_{D_n(\beta)}d\tau\,
\widetilde W_L(\beta-\tau_1,\omega)\widetilde R_L(\tau_1-\tau_2,\omega)\\
&& ...\widetilde R _L(\tau_{n-1}-\tau_n,\omega)\widetilde R_L
(\tau_n,\omega)+{\cal \widetilde R}^{(1)}_{N+1,L}(\beta,\omega).
\end{eqnarray*}
with 
\begin{align} \nonumber 
{\cal \widetilde R}^{(1)}_{N+1,L}(\beta,\omega)=   (-1)^{N+1}\int_{D_{N+1}(\beta)}
d\tau 
W_L(\beta-\tau_1,\omega) \cdot{{ \widetilde R}_L}(\tau_1,\omega)... {{\widetilde R}_L}(\tau_{N+1},\omega) 
 \\\label{refintermed1}
\end{align}
 where $D_n(\beta)$ is defined in \eqref{Dn}. Recall
 that $\widetilde R_L=\delta\omega \widetilde R_{1,L}+(\delta\omega)^2 \widetilde R_{2,L}$. So  
  \eqref{refintermed1} gives  \eqref{restreg1} and   a simple
  induction argument finishes the proof of the lemma. \qed

\vspace{0.5cm}

\bigskip

\noindent {\it Continuing the proof of Theorem \ref{thm3}.} From Theorem \ref{thm2}, 
 we know that for   $\x \in \Lambda_L$,
 and $\beta>0$, $\C\ni \omega \to G_L(\x,\x; \beta, \omega)$ is an
 entire function.

In order to  prove \eqref{deriveenn}, we will show that for all $\x \in \Lambda_L$ we have:
\begin{align} \nonumber
G_L(\x,\x;\beta,\omega)&=G_L(\x,\x;\beta,\omega_0)+ \sum_{n=1}^N (\delta\omega)^n\sum_{k=1}^n 
W_{k,L}^{n-k}(\x;\beta,\omega_0)  \\ \label{cqfm}
&+ {\cal\widetilde  R}_{N+1,L}(\x;\beta,\omega,\omega_0),
\end{align}
 where the remainder term  satisfies the property that its first $N$
 derivatives at $\omega_0$ are zero.
 
 By rewriting  Lemma \ref{lm3} in terms of the corresponding integral kernels, and
 looking at the diagonal of these kernels, we have (remember
that $\phi(\x,\x)=0$):
\begin{eqnarray}\nonumber 
G_L(\x,\x;\beta,\omega)= G_L(\x,\x;\beta,\omega_0) + \sum_{n=1}^{N} (\delta\omega)^{n}\,
W_{n,L}(\x,\x;\beta,\omega,\omega_0)  \\ +{ \cal 
 \widetilde R}^{(1)}_{N+1,L}(\x,\x;\beta,\omega,\omega_0) + { \cal  \widetilde
R}^{(2)}_{N+1,L}(\x,\x;\beta,\omega,\omega_0).&\label{pseudoWL}
\end{eqnarray}
By construction, the two remainders are smooth functions which remain
smooth even if they are divided by $(\delta\omega)^{N+1}$, see
formulas \eqref{restreg1} and \eqref{restreg2}. This means
that their first $N$ derivatives at $\omega_0$ are all zero. Thus the
$N$th derivative of $G_L(\x,\x;\beta,\cdot)$ at $\omega_0$ can only
come from the $W$'s.  

What we still have to do is to remove the $\omega$ dependence from
$W$'s. For, let us show that for $1\leq n\leq N $,  $\x\in\Lambda_L$ and  $\vert \delta\omega\vert < 1$,
\begin{eqnarray}\label{dev-expon44}
W_{n,L}(\x,\x;\beta,\omega,\omega_0)=\sum_{m=0}^{N}\,(\delta\omega)^m\,W_{n,L}^m(\x;\beta,\omega_0) + 
{ {\cal \widetilde R}^{(3)}_{n,N+1}(\x; \beta,\omega,\omega_0) },
\end{eqnarray}
where $W_{n,L}^m$ were introduced in \eqref{def-WL-n-k}, and 
${ {\cal \widetilde R}^{(3)}_{n,N+1}(\x; \beta,\omega,\omega_0) }$ has
its first $N$ derivatives at $\omega_0$ equal to $0$. Indeed, if we
replace the integral kernel of $\widetilde{I}_{k,L}$ from
\eqref{InL2-def} in the expression of $W_{N,L}$ from \eqref{def-WLN},
we see that we can add up all the magnetic phases, and obtain a factor
of the type:
$$\exp\{\phi(\x,\y_1)+\phi(\y_1,\y_2)+\dots +\phi(y_{k-1},\y_k)+
\phi(y_{k},\x)\}.$$
Then this exponent will equal the magnetic flux defined in
\eqref{defFL}, plus an additional contribution $\phi(\x,\x)$ {\it which
  is zero due to the antisymmetry of the magnetic phase}.  Now if we
expand $e^{i(\delta\omega)\Fl_j(\x,...,\y_j)}$ in Taylor series up to the
$N$th order, we obtain \eqref{dev-expon44} where the remainder has
again the property that its first $N$ derivatives at $\omega_0$ are
zero. Now introduce \eqref{dev-expon44} in \eqref{pseudoWL}, and after
some algebra involving the multiplication of two series, we eventually get \eqref{cqfm}. Then we
can identify the $N$th derivative at $\omega_0$ of the kernel's diagonal as the
coefficient multiplying the $N$th power of $\delta\omega$.  The
identity \eqref{deriveenn} is proved. 

Now let us prove the second part of the theorem, i.e. the estimate
\eqref{upperbo}, which is also linked to the natural question "why is formula \eqref{deriveenn} better than
the one from \eqref{expker}"? The answer is that
$W_{k,L}^m(\x;\beta,\omega_0)$ {\it does not grow with} $L$, and we
will see in the next section that it even converges when $L$ tends to
infinity. Let us show here its uniform boundedness in $L$. 

Looking at its definition given in \eqref{def-WL-n-k}, and using the
estimates from \eqref{est_R1L-R2L-RL} together with the diamagnetic
inequality, we see that we need to estimate 
\begin{align}\nonumber
&
|W_{k,L}^m(\x;\beta,\omega)|\leq \mathcal{C}_3^k\sum_{j=1}^k(-1)^j\sum_{(i_1,...,i_j)\in\{1,2\}^{j}}\chi_{j}^{k}(i_1,...,i_j)
\int_{D_j(\beta)}d\tau\int_{\Lambda^j}d\y\\ \nonumber
& 
\frac{|\Fl_{j}(\x,\y_1,...,\y_{j})|^m}{m!}\,\,G_\infty(\x,\y_1;16(\beta-\tau_1))
G_\infty(\y_1,\y_2;16(\tau_1-\tau_2))\\
\label{def-WL-n-k2}
& 
\,...\,G_\infty(\y_{j-1},\y_j;16(\tau_{j-1}-\tau_j))\; G_\infty(\y_j,\x;16\tau_j)
\end{align}

Let  $\alpha= 16 m$, and identify $\x=\y_0$. In view of \eqref{estFL} and the explicit 
form \eqref{heatk} of the heat kernel, for $1\leq l \leq j-1, 1\leq l'\leq l$; $\y_0,\y_1,...,\y_j \in \Lambda_L^{j+1}$ 
and $\{\tau_1, \tau_2,... \tau_j \} \in D_j(\beta)$ we need  the
straightforward estimate 
 \begin{eqnarray}\nonumber 
&\vert\y_{l'-1}-\y_{l'}\vert \vert\y_{l}-\y_{l+1}\vert 
\exp(- \frac {\vert\x-\y_1\vert^2}{2\alpha(\beta-\tau_1)})
...\exp( -\frac {\vert\y_k-\x\vert^2}{2\alpha \tau_k})  \\
&\leq 2 \alpha \beta  \exp( -\frac {\vert\x-\y_1\vert^2}{4\alpha(\beta-\tau_1)})
...\exp( -\frac {\vert\y_k-\x\vert^2}{4\alpha \tau_k}). \label{inter1}
\end{eqnarray}
Thus   \eqref{estFL} and  \eqref{inter1} imply 
\begin{align} \label{estimatFL}
&\vert \Fl_{j}(\x,\y_1,...,\y_{j})\vert^{m}  G_\infty(\x,\y_1;16(\beta-\tau_1))
...
G_\infty(\y_j,\x;16\tau_j) \nonumber \\
&\leq (\sum_{l=1}^{j-1}l\alpha \beta)^{m} 2^{3(j+1)/2} G_\infty(\x,\y_1;32(\beta-\tau_1))
...
G_\infty(\y_j,\x;32\tau_j) \nonumber \\
&\leq 
(8 j^{2}m \beta)^m 2^{3(j+1)/2 } G_\infty(\x,\y_1;32(\beta-\tau_1))
...
G_\infty(\y_j,\x;32\tau_j). 
\end{align}
Integrating over the spatial coordinates, using the semigroup
property \eqref{estnoy}, and then integrating over $\tau$ variables, one eventually
obtains the uniform upper bound in $L$ given in \eqref{upperbo}.

\qed. 

\vspace{0.5cm}

\begin{remark} Theorem \ref{thm3} gives us  what we need for the purpose of this article.  One can show that 
  our  analysis  can be applied in order to get  the off-diagonal
  terms  of the integral kernel i.e. $\frac{\partial^nG_L}{\partial\omega^n}(\x,\x';\beta,\omega)$,
 $(\x,\x') \in \Lambda_L\times \Lambda_L$ and $\beta>0, \omega>0$. In that case we get 
\end{remark} 
\begin{align}\nonumber 
&\frac{1}{n!}\,\frac{\partial^nG_L}{\partial\omega^n}(\x,\x';\beta,\omega)=
\frac{i^n\phi^n(\x,\x')}{n!}G_L(\x,\x';\beta,\omega)\\
& + 
\sum_{k=1}^n\sum_{j=1}^k(-1)^j\int_{D_j(\beta)}d\tau 
\sum_{(i_1,...,i_j)\in\{1,2\}^{j}}\chi_{j}^{k}(i_1,...,i_j)
\int_{\Lambda^j}d\y \nonumber  \\
&\cdot \frac{\left(i(\sum_{l=1}^{j-1}\fl(\x,\y_{l},y_{l+1}) + 
\fl(\x,\y_{j},\x')+ \phi(\x,\x'))\right)^{n-k}}{(n-k)!}\nonumber \\
& 
\cdot G_L(\x,\y_1,\beta-\tau_1,\omega_) R_{i_1,L}(\y_1,\y_2,\tau_1-\tau_2,\omega)
\,...R_{i_j,L}(\y_j,\x',\tau_j,\omega).
\end{align}
\vspace{0.5cm}

\begin{remark} Let $ \beta>0 $ and $\omega_0 \geq 0$ and $\omega \in \C$. From  \eqref{nHS} we have 
$$ \Vert W_L(\beta-\tau, \omega, \omega_0 ) \widetilde R_L(\tau,\omega, \omega_0)\Vert_1 \leq  
\Vert W_L(\beta-\tau, \omega, \omega_0 )\Vert_2  \Vert\widetilde R_L(\tau,\omega, \omega_0)\Vert_2  $$
$$ \leq \frac {C(\beta,\omega,  \omega_0)L^3}{(\beta-\tau)^{3/4}\tau^{3/4}}, $$
where  $C(\beta, \omega, \omega_0)=\cst_3(\beta,\omega_0)
(\mid \delta\omega \mid +\mid (\delta\omega) \mid^2)$ and $\cst_3(\beta,\omega_0)$ 
is given in \eqref{est_R1L-R2L-RL}. Then the $B_1$-operator
valued function $ \tau \in (0, \beta) \to W_L(\beta-\tau, \omega ) \widetilde R_L(\tau,\omega, \omega_0)$
 is $B_1$-integrable. Denote  by $ U_L(\beta,\omega, \omega_0)=
\int_0^\beta W_L(\beta-\tau, \omega ) \widetilde R_L(\tau,\omega, \omega_0)$,
 \begin{equation} \label{normU}
 \Vert U_L(\beta,\omega, \omega_0) \Vert_1 \leq 
 C L^3\int_0^\beta  \frac {1}{(\beta-\tau)^{3/4}\tau^{3/4}} \leq \frac{16 C L^3} {\sqrt \beta}.
\end{equation}
The Duhamel-type formula \eqref{perturb-semi-groupe-bis} then implies that 
$ \widetilde W_L(\beta, \omega )$ is of trace class
 as   a sum of  the two trace class operators, $W_L$ and $U_L$.  Consequently   the operators 
${\widetilde I_{k,L}}(i_1,...,i_k)(\beta,\omega,\omega_0)$ 
defined in \eqref{InL2-def} are of trace class because the integrals
only involve $B_1$-integrable functions. 
\end{remark}

%

\section{Large volume  behavior}
\setcounter{equation}{0}
For further applications in Section 4 we need  to have 
a similar result as in Theorem \ref{thm3} but with  $L= \infty$. The results of Section 2 cannot 
  be applied to  this situation.  On the contrary, we will show in this section that Theorem \ref{thm3}
remains true even if we take $L= \infty$, and the quantities at finite
volume converge pointwise to the ones defined on the whole space.
 
Recall first  that  the explicit form of the integral
kernel of $e^{-\beta H_\infty ( \omega)}; \beta >0, \omega\geq 0$ is given by 
\begin{align} \label{intkinfini}
& G_\infty(\x,\x'; \beta, \omega)= \frac {1}{(2\pi\beta)^{3/2}} 
\frac {\omega\beta/2}{\sinh(\omega\beta/2)} 
e^{i \omega \phi(\x,\x')}  \\  
&\cdot \exp {\left\{-\frac {1}{2\beta}\left(\frac {\omega\beta/2}{\tanh( \omega\beta/2)}[ (x_1-x'_1)^2 
+  (x_2-x'_2)^2 ] + (x_3-x'_3)^2\right)\right \}}\nonumber 
\end{align}
where  the phase $\phi$ is defined in \eqref{phaseflux}.

We start with a technical result. For any $\x \in \Lambda_L$, we
denote with $d(\x) := \rm{dist}(\x, \partial \Lambda_L)$. Let  
 $ M:=\{ (\x,\x') \in \Lambda_L\times \Lambda_L $: 
$d(\x)\leq 1 $ or $d(\x')\leq 1 \}$, and denote with  $\chi_{M}$ 
the characteristic function of ${ M}$.


\begin {theorem} \label{thm31}
Let $\beta>0$ and  $\omega\geq 0$. Then for any  
$(\x,\x')\in \Lambda_L \times \Lambda_L$ we have:
\begin{align} \label{estdiffoncgr1}
 & \vert  G_L(\x,\x';\beta, \omega)-  G_\infty (\x,\x';\beta, \omega) \vert \leq 
 2 \chi_{ M} (\x,\x')G_\infty(\x,\x';\beta)\\ 
&+ \nonumber \cst_4 
(1- \chi_{ M})(\x,\x')  G_\infty (\x,\x';16\beta)
  e^ {-(\frac{d^2(\x)}{64\beta}
 +\frac{d^2(\x')}{64 \beta})},
\end{align}
and 
\begin{align} \label{estdiffoncgr2}
&\vert (-i\nabla_{\x} -\omega \A(\x)) [ G_L(\x,\x';\beta, \omega)- 
 G_\infty (\x,\x';\beta, \omega)]\vert \\ 
& \leq \nonumber 
  \frac{  \cst_5}{\sqrt \beta}  \chi_M(\x,\x') G_\infty(\x,\x';
 8\beta) + \cst_6 (1- \chi_M) G_\infty (\x,\x';16\beta) 
e^ {-(\frac{d^2(\x)}{64\beta}
 +\frac{d^2(\x')}{64 \beta})}. 
\end{align}
 where $\cst_4=\cst_4(\beta,\omega)= c(1+  \beta)^6(1+ \omega)^4  $, $\cst_5=
 \cst_5(\beta,\omega) =c(1+  \beta)^5(1+ \omega)^3 $, 
$\cst_6=\cst_6(\beta,\omega)= c (1+\beta)^8 (1+\omega)^5 $ and $c>1$
 is  a numerical constant.
\end{theorem}
To prove the theorem, we need the following lemma.
\begin{lemma} \label{lm4}
Let $\beta>0$, $\omega\geq 0$ and $\alpha \in \{0,1\}$. Then for every
$(\x,\x')\in \Lambda_L \times \Lambda_L$ we have:
\begin{align} \label{diffoncgr}
& \partial^\alpha _{x_i} G_L(\x,\x';\beta, \omega)-
\partial_{x_i}^\alpha  G_\infty (\x,\x';\beta, \omega)
\\ 
& \nonumber \frac{1}{2}\int_0^\beta d\tau \int_{\partial \Lambda_L} 
d \sigma(\y)\;  \partial_{x_i}^\alpha  G_\infty (\x,\y;\tau, \omega)[{\bf n}_\y\cdot \nabla_{\y} G_L(\y,\x';\beta-\tau, \omega)],
\end{align}
where $d \sigma(\y) $ is the  measure on $\partial \Lambda_L$ and
${\bf n}_\y$ is the outer normal to $\partial \Lambda_L$ at $\y$.
\end{lemma}
\noindent {\it Proof.} Let $\beta>0$, $\omega \geq 0$. 
Recall that both Green functions 
$G_\infty (\x,\x';\tau, \omega)$ 
and  $ G_L (\x,\x';\tau, \omega)$ satisfy in
$\Lambda_L\times\Lambda_L$ in distributional sense the equation 
\begin{align} \label{equadiff}
 & {\rm (i)} \quad \partial_\tau G(\x,\x';\tau ) 
= -\frac{1}{2}[-i\nabla_{\x} -\omega \A(\x)]^2G(\x,\x';\tau ); 
\quad \tau >0 \\ \label{condinit1}
& {\rm (ii)} \quad G(\x,\x'; \tau=0_+)= \delta(\x-\x').
\end{align}
For $0<\tau< \beta $  and on 
$\Lambda_L \times \Lambda_L$, define  the following quantity:
\begin{equation} \label{defQ}
Q(\x,\x';\beta,\tau):= \int_{\Lambda_L}d\y\overline{G_L (\y,\x; \beta-\tau, \omega)}G_\infty (\y,\x';\tau, \omega).
\end{equation}
Then by  \eqref{equadiff} it is easy to see that
\begin{align}  \label{derQ1}
 &\partial_\tau Q(\x,\x';\beta,\tau)\\ &= \frac{1}{2} \int_{\Lambda_L}d\y
\left \{\overline{(-i\nabla_{\y} -\omega \A(\y))^2G_L (\y,\x;\beta-\tau,
    \omega)}G_\infty (\y,\x';\tau, \omega )\frac{}{}\right .\nonumber \\
&\left .\frac{}{}- 
\overline{G_L (\y,\x;\beta-\tau, \omega)}(-i\nabla_{\y} -\omega
\A(\y))^2G_\infty (\y,\x';\tau, \omega)\right \}.\nonumber
\end{align}
Since $G_L (\x,\x';\tau, \omega)=0$ if $\x \in \partial\Lambda_L$ or $\x'\in \partial\Lambda_L$, integration by parts
gives:

\begin{align}  \label{derQ2}
\partial_\tau Q(\x,\x';\beta,\tau)= -\frac{1}{2} \int_{\partial \Lambda_L}d\sigma(\y)
\overline { {\bf n}_\y \cdot \nabla_{\y} G_L(\y,\x;\beta-\tau, \omega)}
  G_\infty (\y,\x';\tau, \omega).
\end{align}
Now by integrating with respect to $\tau$ from $0_+$ to $\beta_-$, and
using \eqref{condinit1}, we obtain:
\begin{align} \label{diffoncgr1}
& G_L(\x,\x';\beta, \omega)- G_\infty (\x,\x';\beta, \omega)
\\ &= \nonumber \frac{1}{2}\int_0^\beta d\tau \int_{\partial
  \Lambda_L} d\sigma(\y)  \overline {{\bf n}_\y\cdot \nabla_{\y} G_L(\y,\x;\beta-\tau, \omega)}
  G_\infty (\y,\x';\tau, \omega).
\end{align}
Now using the self-adjointness property of the semigroup we obtain
$G(\x,\y;\tau) =\overline {G(\y,\x;\tau) }$, thus we can rewrite \eqref{diffoncgr1} as: 
\begin{align} \label{diffoncgr2}
& G_L(\x,\x';\beta, \omega)- G_\infty (\x,\x';\beta, \omega) \\ 
&=\nonumber \frac{1}{2}\int_0^\beta d\tau \int_{\partial \Lambda_L} 
d\sigma(\y)\; G_\infty (\x,\y;\tau, \omega)[{\bf n}_\y \cdot \nabla_{\y} G_L(\y,\x';\beta-\tau, \omega)].
\end{align}
The lemma now follows from \eqref{diffoncgr2}. \qed 

\vspace{0.5cm}

\noindent {\it Proof of Theorem \ref{thm31}.} Let  $\beta>0$,  $\omega \geq 0$ and suppose first that  $(\x,\x') \in  M $.
  Then \eqref{estdiffoncgr1} follows
from the diamagnetic inequality \eqref{diam}. Let us show \eqref{estdiffoncgr2} in the same case.  
We know from \eqref{estdG} that 
\begin{equation} \label{thm311}
\vert (-i\nabla_{\x}-\omega \A(\x)) G_L(\x,\x';\beta, \omega)\vert \leq 
\frac{ \cst }{\sqrt \beta} \, G_\infty(\x,\x';
8\beta).
\end{equation}
On the other hand, using the observation that $-i\nabla_{\x}-\omega
\A(\x)$ is transformed into $-i\nabla_{\x}-\omega \A(\x-\x')$ after
commutation with $e^{i\phi(\x,\x')}$, then by direct computation from \eqref{intkinfini} we
get that for  all $\eta >0$, 
\begin{align} \label{thm312}
\vert  (-i\nabla_{\x}-\omega \A(\x)) G_\infty(\x,\x'; \eta \beta, \omega)\vert \leq 
\frac{ \cst'_1}{ \sqrt{\eta \beta}} G_\infty(\x,\x';
2\eta\beta),
\end{align}
where $\cst'_1=\cst'_1(\beta, \omega)=  2(1+ \eta)(1+  \omega) (1+\beta)$. Then \eqref{thm311} and  \eqref{thm312} for $\eta= 1\medskip
$ imply \eqref{estdiffoncgr2}.

Now suppose that $(\x,\x') \not\in M$. This means that neither points are near the boundary. For  $\y \in
\partial \Lambda$ then by \eqref{estdG} we have:
\begin{equation} \label{thm313}
\vert \nabla_{\y} G_L(\y,\x;\beta-\tau, \omega)\vert \leq 
\frac{ \cst_1}{\sqrt {\beta- \tau} }\,G_\infty(\y,\x,
8(\beta-\tau)).
\end{equation}

By applying  the estimates  \eqref{thm313}, \eqref{diam} and  the Lemma \ref{lm4}, we get

\begin{align} \label{thm314}
 &\vert G_L(\x,\x';\beta, \omega)-  G_\infty (\x,\x';\beta, \omega) \vert  
 \\ & \leq \nonumber  2^{7/2}\cst_1\int_0^\beta d\tau \int_{\partial \Lambda_L} d\sigma(\y)
\frac{ G_\infty(\y,\x;
8(\beta-\tau))}{\sqrt {\beta- \tau} } 
    G_\infty (\y,\x';8\tau).
\end{align}
But  if  $ \y \in \partial \Lambda_L $, $ \vert \x-\y \vert \geq d(\x)$, then  a straightforward estimate 
 shows that  for $0<t< \beta $, we have $  G_\infty (\y,\x; 8t) \leq e^ {-\frac{d^2(x)}{32\beta}}G_\infty (\y,\x;16 t)$.
Thus we get 
\begin{align} \label{thm315}
 &\rm{r.h.s.   \: of \: \eqref{thm314}} \\ 
&\leq
  2^5\cst_1 \int_0^\beta d\tau  \frac{e^ {-(\frac{d^2(x)}{32(\beta-\tau)} +\frac{d^2(x')}{32\tau})}}
{\sqrt {\beta- \tau}}
\int_{\partial \Lambda_L} d\sigma(\y) G_\infty(\y,\x;
16(\beta-\tau))
    G_\infty (\y,\x';16\tau).\nonumber 
\end{align}
For any $t,t' >0$, let us look at the integral 
\begin{equation}\label {thm3151} 
\int_{\partial \Lambda_L} d\sigma(\y) 
 G_\infty(\y,\x;
t)G_\infty (\y,\x';t').
\end{equation}
Using the convexity of $\Lambda_L$, replacing the integrals on the
sides of $\partial \Lambda_L$ by integrals on $\R^2$ (thus getting an
upper bound), and using the semigroup property in two dimensions, we
can show that there exists a numerical constant $C>0$ such that 
  \begin{equation}\label {thm31513} 
\int_{\partial \Lambda_L} d\sigma(\y) 
 G_\infty(\y,\x;
t)G_\infty (\y,\x';t')\leq C \;\frac{\sqrt{t+t'}}{\sqrt{t}\; \sqrt{t'}}G_\infty (\x,\x';t+t').
\end{equation}
To be more precise, let us look at the integral on the hyperplane
defined by $H:=\R^2+(L/2,0,0)$:
 \begin{equation}\label {thm315134} 
\int_{H} d\sigma(\y) 
 G_\infty(\y,\x;
t)G_\infty (\y,\x';t'),
\end{equation}
where $\x$ and $\x'$ are on the same side of $\R^3$ with respect to $H$. Decompose
$\x=\x_1+\x_2$ and $\x'=\x_1'+\x_2'$ where $\x_1$ and $\x_1'$ are the
parallel components with $H$, while $\x_2$ and $\x_2'$ are the
orthogonal components on $H$. Note that here $|\x_2|^2+|\x_2'|^2\geq
|\x_2-\x_2'|^2$. Since $|\x-\y|^2=|\x_1-\y|^2+|\x_2|^2$ if $\y\in H$,
we can explicitly integrate with respect to $\y$ and eventually get
\eqref{thm31513}. 

Then we can write:
$$\int_{\partial \Lambda_L} d\sigma(\y) 
 G_\infty(\y,\x;
16(\beta-\tau))G_\infty (\y,\x';16\tau) \leq
{\cst_1'}\frac{\sqrt{\beta}}{\sqrt{(\beta-\tau)\tau}} G_\infty(\x,\x';
16\beta).$$
Therefore  since $\x,\x'$ satisfy $d(\x),d(\x') \geq 1 $ we get 
\begin{align} \label{thm316}
 & \vert G_L(\x,\x';\beta, \omega)-  G_\infty (\x,\x';\beta, \omega) \vert 
 \leq  2^4 \cst_1\cst_1' G_\infty (\x,\x';16\beta)
 \\ \nonumber & \cdot e^ {-(\frac{d^2(\x)}{64\beta}
 +\frac{d^2(\x')}{64 \beta})}\int_0^\beta d\tau \frac{e^ {-(\frac{1}{64(\beta-\tau)}
 +\frac{1}{64\tau})}}
{\sqrt { \beta- \tau} }\frac{\sqrt{\beta}}{\sqrt{(\beta-\tau)\tau}}.
\end{align}
Due the exponential decay, there are no singularities in this integral,
and a straightforward estimate gives \eqref{estdiffoncgr1}. 

We now use the same method as above to 
prove \eqref{estdiffoncgr2} in the case when $(\x,\x')\not\in M$.  We 
know  from Lemma \ref{lm4} that
\begin{align} \label{thm318}
 & (-i\nabla_{\x}-\omega \A(x)) \left( G_L(\x,\x';\beta, \omega)-  G_\infty (\x,\x';\beta, \omega) \right)
 =  \frac{1}{2}\int_0^\beta d\tau \int_{\partial
   \Lambda_L} d\sigma(\y)\nonumber\\ &  \cdot 
(-i\nabla_{\x}-\omega \A(x))   G_\infty (\x,\y;\tau, \omega)
 [{\bf n}_\y \cdot \nabla_{\y} G_L(\y,\x';\beta-\tau, \omega)].
\end{align}
Then  by \eqref{thm311}  and \eqref{thm312} 
\begin{eqnarray} \label{thm319}
 \vert (-i\nabla_{\x}-\omega \A(x)) \left( G_L(\x,\x';\beta, \omega)-  
G_\infty (\x,\x';\beta, \omega) \right) \vert \leq 
 \\ \nonumber  4\cst_1 \cst'_1\int_0^\beta d\tau 
\int_{\partial \Lambda_L} d\sigma(\y)  \frac{G_\infty (\y,\x;8(\beta-\tau))}{\sqrt{\beta-\tau}}
  \frac{ G_\infty (\y,\x';8\tau)}{\sqrt \tau }.
\end{eqnarray}
Then by using   the same arguments leading to \eqref{thm316} we get
\begin{align} \nonumber
 &\vert (-i\nabla_{\x}-\omega \A(x)) \left( G_L(\x,\x';\beta, \omega)-  
G_\infty (\x,\x';\beta, \omega) \right) \vert 
 \\ \label{thm320} &\leq 16\sqrt{\beta}\cst_1 {\cst'_1}^2 G_\infty (\x,\x';16\beta)e^ {-(\frac{d^2(x)}{64\beta}
 +\frac{d^2(x')}{64 \beta})}\int_0^\beta d\tau \frac{e^ {-(\frac{1}{64(\beta-\tau)}
 +\frac{1}{64\tau})}}{\tau (\beta- \tau)},
\end{align}
from which \eqref{estdiffoncgr2} follows. Theorem \ref{thm31} is
proved. \qed
  
\vspace{0.5cm}

We now want to prove that the equality \eqref{deriveenn} stated in
Theorem \ref{thm3} remains true even if $L$ tends to infinity. 
It is well known  (see e.g.\cite{AC}) that for $\beta >0, \omega\geq 0$ and $(\x,\x') \in \R^3\times \R^3$, 
\begin{equation}\label{konverg1}
G_{\infty}(\x,\x';\beta,\omega)=
\lim_{L\rightarrow\infty}G_L (\x,\x';\beta,\omega).
\end{equation}
Our main goal now is to show that this pointwise convergence holds
true for all the derivatives 
 $\frac{\partial^n
  G_{L}}{\partial\omega^n}$, $n \geq 1$. 

\medskip

We need to introduce some notation. Let $\beta>0 $ and $\omega \geq 0 $. 
For $(\x,\x') \in \R^3\times \R^3$ define 
\begin{align}\label{def_R1-R2-R1infini}
R_{1,\infty}(\x,\x';\beta,\omega)& := \A(\x-\x')\cdot
\left(i\nabla_\x+\omega
  \A\left(\x\right)\right)G_\infty(\x,\x';\beta,\omega).\nonumber \\
 R_{2, \infty}(\x,\x';\beta,\omega)& :=  \frac{1}{2}\,
\A^2(\x-\x')G_\infty(\x,\x';\beta,\omega).
\end{align}
Le us note that we again have the same type of estimates as in
\eqref{est_R1L-R2L-RL0}, \eqref{est_R1L-R2L-RL1} and
\eqref{est_R1L-R2L-RL}, i.e. gaussian localization in the
difference of the spatial arguments. The linear growth of the magnetic
potential disappears when one commutes $-i\nabla_\x$ with the magnetic
phase, as we have already seen in \eqref{thm312}. 

Now define for $\x\in\R^3$, $k\geq 1, m\geq 0$: 
\begin{align}\nonumber
&
W_{k,\infty}^m(\x;\beta,\omega):=\sum_{j=1}^k(-1)^j\sum_{(i_1,...,i_j)\in\{1,2\}^{j}}\chi_{j}^{k}(i_1,...,i_j)
\int_{D_j(\beta)}d\tau\int_{\R^{3j}}d\y\\ \nonumber
& 
\frac{\left(i\left(\Fl_{j}(\x,\y_1,...,\y_{j})\right)\right)^m}{m!}\,\,G_\infty(\x,\y_1;\beta-\tau_1,\omega)
R_{i_1,\infty}(\y_1,\y_2;\tau_1-\tau_2,\omega)\\
\label{def-Winfty-n-k}
& 
\,...\,R_{i_{j-1},\infty}
(\y_{j-1},\y_k;\tau_{j-1}-\tau_j,\omega)R_{i_j,\infty}(\y_j,\x;\tau_j,\omega).
\end{align}
Since every integrand is bounded by a free heat kernel, and because
the flux $Fl_j$ can be bound by {\it differences} of its arguments
(see \eqref{estFL}), then the above multiple integrals are absolutely
convergent. Also note the important thing that multiplication by
$|\y-\y'|^m$ of the free heat kernel only improves the singularity in
the time variable due to the estimate 
\begin{equation}\label{kahas2}
|\y-\y'|^m \e^{-|\y-\y'|^2/t}\leq {\rm const}\cdot
t^{m/2}\e^{-|\y-\y'|^2/(2t)}.
\end{equation}
The last important remark about $W_{k,\infty}^m(\x;\beta,\omega)$ is
that {\it it does not depend on} $\x$. This can be seen by factorizing
all the magnetic phases which enter in the various factors of the
integrand, and see that they add up to give another $Fl_j$, which only
depends on differences of variables. The remaining factors are also
just functions of differences of variables. Therefore by changing $\x$
we get the same value for $W_{k,\infty}^m$ after a change of variables
(a translation) in all integrals, since $W_{k,\infty}^m$ only involves
integrals defined on the whole space.

Then we have 
\begin{theorem}\label{thm32}
Let $\beta>0$ and  $\omega\geq 0$. Fix $ \x\in\R^3$ and $n \geq
1$. Then we have:
\begin{equation}\label{lim-des-noyaux-diag}
\frac{1}{n!}\frac{\partial^n
  G_{\infty}}{\partial\omega^n}\,(\x,\x;\beta,\omega)=
\lim_{L\rightarrow\infty}\,\frac{1}{n!}\frac{\partial^{n}G_L}{\partial\omega^{n}}(\x,\x;\beta,\omega)=
\sum_{k=1}^{n} W_{k, \infty}^{n-k}(\x;\beta,\omega).
\end{equation}
\end{theorem}
\proof  Fix $ \beta>0$ and $\omega\geq 0$. Let $n\geq 1$ and
$(\x,\x') \in \R^3$. Choose $L$ large enough such that $\x\in
\Lambda_L$. Then from
 \eqref{estdiffoncgr1} and \eqref{estdiffoncgr2} we have: 
\begin{align} 
&\vert  G_L(\x,\x';\beta, \omega)-  G_\infty (\x,\x';\beta, \omega) \vert \leq \cst_4  \beta^{-3/2}
    e^ {-(\frac{d^2(\x)}{64\beta} +\frac{d^2(\x')}{64 \beta})},\label{thm40} \\\nonumber
&\vert R_{1,L}(\x,\x';\beta,\omega)-R_{1,\infty}(\x,\x';\beta,\omega) \vert \leq 
\cst_6 \beta ^{-1}  e^ {-(\frac{d^2(\x)}{64\beta} +\frac{d^2(\x')}{64 \tau})}, \\\nonumber
&\vert R_{2,L}(\x,\x';\beta,\omega)-R_{2, \infty}(\x,\x';\beta,\omega)\vert \leq 
\cst_4 \beta^{-1/2}
    e^ {-(\frac{d^2(\x)}{64\beta} +\frac{d^2(\x')}{64 \beta})}.
\end{align}
Then  for all $(\x,\x') \in \R^3\times \R^3$,  estimates \eqref{thm40} show  respectively that  
$$\lim_{L\to\infty}G_L(\x,\x',\beta,\omega)
 =  G_{\infty}(\x,\x',\beta,\omega),$$  
$$\lim_{L\rightarrow\infty}R_{i,L}(\x,\x',\beta,\omega) =  R_{i,\infty}(\x,\x',\beta,\omega),$$
for $i=1,2$. Then 
$$\lim_{L \to \infty}G_L(\x,\y_1;\beta-\tau_1,\omega)
R_{i_1,L}(\y_1,\y_2;\tau_1-\tau_2,\omega)...
R_{i_j,L}(\y_j,\x;\tau_j,\omega) =$$
$$G_\infty(\x,\y_1;\beta-\tau_1,\omega)
R_{i_1,\infty}(\y_1,\y_2;\tau_1-\tau_2,\omega)...
R_{i_j,L}(\y_j,\x;\tau_j,\omega).$$
Furthermore  by \eqref{est_R1L-R2L-RL} and \eqref{diam}  we have
$$\vert G_L(\x,\y_1;\beta-\tau_1,\omega)
R_{i_1,L}(\y_1,\y_2;\tau_1-\tau_2,\omega)...
R_{i_j,L}(\y_j,\x;\tau_j,\omega) \vert \leq $$
$$ 4^3\cst_3^jG_\infty(\x,\y_1;16(\beta-\tau_1))
G_\infty(\y_1,\y_2;16(\tau_1-\tau_2))...
G_\infty(\y_j,\x;16\tau_j). $$
 this last quantity  is  $L$-independent and $\R^3 $-integrable by the the semigroup property since
$$ \int_{\R^{3j}}d\y G_\infty(\x,\y_1;16(\beta-\tau_1))...
G_\infty(\y_j,\x;16\tau_j) = G_\infty(\x,\y_1;16\beta).$$ 

Note that the flux $Fl_j$ does not influence anything, since it can be
bound by powers of differences between spatial variables, which will
meet the gaussian decay of the free heat kernels. Thus they will 
only affect the time integrals (by making them even less singular).

Then by applying Lebesgue's dominated convergence theorem we get
from   \eqref{def-WL-n-k} and \eqref{deriveenn},
\begin{equation} \label{thm43}
\lim_{L \to \infty}\frac{1}{n!}\,\frac{\partial^nG_L}
{\partial\omega^n}(\x,\x,\beta,\omega)=\sum_{k=1}^n
W_{ k,\infty}^{n-k}(\x;\beta,\omega).
\end{equation}

Now the remaining thing is to show that this also equals $\frac{1}{n!}\,\frac{\partial^nG_\infty}
{\partial\omega^n}(\x,\x,\beta,\omega)$. Fix $ \omega_0 \geq 0$ and
choose $\omega \in\R$ such that
 $\vert \delta\omega \vert = \vert \omega -\omega_0 \vert \leq
 1$. From the usual Taylor formula we can write
\begin{align}\label{tailorform}
G_L(\x,\x,\beta,\omega)&=\sum_{n=0}^N (\delta\omega)^n\frac{1}{n!}\,\frac{\partial^nG_L}
{\partial\omega^n}(\x,\x,\beta,\omega_0)\nonumber \\
&+(\delta\omega)^{N+1}\frac{1}{(N+1)!}\,\frac{\partial^{N+1}G_L}
{\partial\omega^{N+1}}(\x,\x,\beta,\omega_1),
\end{align} 
where $\omega_1$ is between $\omega_0$ and $\omega$. Then by taking
$L$ to infinity, we easily get the estimate (note that $\sum_{k=1}^{N+1}W_{
  k,L}^{N+1-k}(\x;\beta,\omega_1)$ is bounded by a constant
independent of $L$, see the estimate from \eqref{upperbo}) :
\begin{align}\label{tailorform2}
\left \vert   G_\infty(\x,\x,\beta,\omega)-G_\infty(\x,\x,\beta,\omega_0)-
\sum_{n=1}^N (\delta\omega)^n\sum_{k=1}^nW_{
  k,\infty}^{n-k}(\x;\beta,\omega_0) \right \vert \nonumber \\
\leq C(N,\beta)|\delta\omega|^{N+1}.
\end{align} 
Since $G_\infty(\x,\x;\beta,\omega)$ is smooth in $\omega$ (see
\eqref{Ginfinidiag}), it follows that the coefficient of
$(\delta\omega)^n$ must equal $\frac{1}{n!}\,\frac{\partial^nG_\infty}
{\partial\omega^n}(\x,\x,\beta,\omega_0)$, and we are done. \qed 

\vspace{0.5cm}

%
\section{ Thermodynamic limit for  magnetic susceptibilities}
\setcounter{equation}{0}
As a consequence of the analysis of the previous section we are now able to prove 
the main technical result of this paper:

\begin{theorem}\label{thm41}
 Let $ n \geq 1$, $\beta > 0$ and  $\omega\geq 0$.   
\begin{eqnarray}\label{majodiff}
\int_{\Lambda_L}d\x\, \left | \frac{\partial^n G_\infty}{\partial\omega^n}(\x,\x;\beta,\omega)
- \frac{\partial^n G_L}{\partial\omega^n}(\x,\x;\beta,\omega)
\right|\leq\,L^2\cst(n,\beta, \omega).
\end{eqnarray}
where $ \cst(n,\beta,\omega):= c(n)\frac{(1+ \beta)^{7n+8}}{\sqrt \beta}(1+ \omega)^{3n+5}$ where
$c(n)$  only depends on $n$.
\end{theorem}

\noindent {\it Proof.} Let $ 1 \leq j \leq k$, $1 \leq k \leq n$,  $1
\leq m \leq n-k$, and denote the integrand in \eqref{def-Winfty-n-k} with:
\begin{align}\label{efjei1}
  F_{j,\infty}^m &=F_{j,\infty}^m(\x,\y_1,...,\y_j,\tau_1,...,\tau_j,\beta,\omega)\\
& =
\frac{(i\Fl_j(\x,\y_1,...,\y_j))^{m}}{m!}
G_{\infty}(\x,\y_1,\beta-\tau_1,\omega)
... R_{i_j,\infty}(\y_j,\x,\tau_j,\omega).\nonumber 
\end{align}
Denote also by:
\begin{align}\label{efjei10}
  F_{j,L}^m &=F_{j,L}^m(\x,\y_1,...,\y_j,\tau_1,...,\tau_j,\beta,\omega)\\
& =
\frac{(i\Fl_j(\x,\y_1,...,\y_j))^{m}}{m!}
G_{L}(\x,\y_1,\beta-\tau_1,\omega)
... R_{i_j,L}(\y_j,\x,\tau_j,\omega).\nonumber 
\end{align}

 Let $ n \geq 1$, $\beta \geq 0$,  $\omega\geq 0$, and fix $\x \in
 \R^3$. Then by applying the Theorem \ref{thm32}, we can split the
 integrals from $W$'s in "inner" and "outer" regions: 
\begin{equation} \label{thm410}
\frac{\partial^n G_{\infty}}{\partial\omega^n}(\x,\x,\beta,\omega)=f_L^n(\x,\beta,\omega)+g_L^n(\x,\beta,\omega)
\end{equation}
where
\begin{eqnarray}\nonumber
f_L^n(\x,\beta,\omega):=n!\sum_{k=1}^{n}\sum_{j=1}^{k}(-1)^{j}
\sum_{(i_1,...i_j)\in\{1,2\}^{j}} \chi_{j}^{k}(i_1,...,i_j)\int_{D_j(\beta)}d\tau\int_{\Lambda_L^j}&d\y
 \\  F_{j,\infty}^{n-k}(\x,\y_1,...,\y_j,\tau_1,...,\tau_j,\beta,\omega), \label{def-fLN}
\end{eqnarray}
\begin{eqnarray} \label{def-gLN}
g_L^n(\x,\beta,\omega):=n!\sum_{k=1}^{n}\sum_{j=1}^{k}(-1)^{j}
\sum_{(i_1,...i_j)\in\{1,2\}^{j}} \chi_{j}^{k}(i_1,...,i_j) \sum_{l=1}^{j}\int_{D_j(\beta)}d\tau \\  \nonumber
 \int_{\R^3}d\y_1
...\int_{\R^{3}\backslash\Lambda_L}d\y_{l}\int_{\R^{3}}d\y_{l+1}...\int_{\R^{3}}d\y_j\,F_{j,\infty}^{n-k}
(\x,\y_1,...,\y_j,\tau_1,...,\tau_j,\beta,\omega).
\end{eqnarray}

Let us now show that 
\begin{eqnarray}\label{thm411}
\int_{\Lambda}d\x\, \left | f_L^n(\x,\beta,\omega)- \frac{\partial^n G_L}{\partial\omega^n}(\x,\x;\beta,\omega)
\right|\leq\,L^2 f(n,\beta, \omega),
\end{eqnarray}
  where $ f(n,\beta,\omega):= c(n)\frac{(1+ \beta)^{7n+3}}{\sqrt \beta}(1+ \omega)^{3n+2}$, 
 $c(n)$ depending  only on $n$.

From now, for the sake of simplicity  we often omit the explicit
dependence of all variables. 
In view of \eqref{def-WL-n-k},\eqref{deriveenn}  and \eqref{def-Winfty-n-k},
 \eqref{lim-des-noyaux-diag}, we  need to estimate
\begin{eqnarray} \nonumber 
F_{j, \infty}^m-F_{j,L}^m=\frac{(i\Fl_j)^{m}}{m!}\{(G_{\infty}-G_L)R_{i_1,\infty}...R_{i_j,\infty}+ 
\\ \label{diffint}  \sum_{l=1}^{j} G_{L}R_{i_1,L}...R_{i_{l-1},L}(R_{i_{l},\infty}-R_{i_{l},L})
 R_{i_{l+1},\infty}...R_{i_j,\infty} \}.
\end{eqnarray}

Denote by  $\chi(\x)$ the characteristic of
$  \{ \x \in \Lambda, d(\x) \leq 1\}$.  Thanks to  the  Theorem \ref{thm31}, we have 
\begin{align} \label{thm412} 
& \vert(R_{1,\infty}-R_{1,L})(\x,\x'; \beta, \omega) \vert \\
& \leq \vert \A(\x-\x') (i\nabla_{\x} +\omega \A)
 \left( G_\infty- 
 G_L \right)(\x,\x';
 \beta, \omega)\vert  \nonumber \\ 
&\leq  \cst_7  G_\infty(\x,\x';
32 \beta) \left(  \chi(\x) + \chi(\x')
  + e^ {-(\frac{d^2(\x)}{64\beta}
 +\frac{d^2(\x')}{64 \beta})} \right),\nonumber 
\end{align}
where $\cst_7 = \cst_7( \beta, \omega)=  c (1+\beta)^9 (1+\omega)^5 $ for some  
 numerical constant $c>1$. But again by the Theorem \ref{thm31} we may  use the  bound

\begin{eqnarray}\label{thm413} 
\vert  (G_\infty-  G_L)(\x,\x';
 \beta, \omega) \vert,
\vert(R_{2,\infty}-R_{2,L})(\x,\x';
 \beta, \omega) \vert  \leq   \\ \nonumber 
 \cst_7  G_\infty(\x,\x';
32 \beta) \left(  \chi(\x) + \chi(\x')
  + e^ {-(\frac{d^2(\x)}{64\beta}
 +\frac{d^2(\x')}{64 \beta})} \right).
\end{eqnarray}
  On the other hand  by  \eqref{est_R1L-R2L-RL},
\eqref{thm312}and   \eqref{def_R1-R2-R1infini},  the kernel of
$R_{i,\infty}$, $ i=1,2$  
 and of $R_{i,L}$, $i=1,2$  satisfy the  inequality 
\begin{equation}  \label{thm4130} \max \left \{\vert(R_{i,\infty}(\x,\x';
 \beta, \omega)  \vert, \;  \vert R_{i,L}(\x,\x';
 \beta, \omega) \vert \right \} \leq  \cst'_3 G_\infty (\x,\x';
 16\beta), 
\end{equation}
  where $ \cst'_3 =  \cst'_3\beta,\omega)= c\cdot \cst_1$, $\cst_1$ is 
defined in \eqref{lm12} and  $c>1$ is a numerical constant which is
chosen large enough such that 
we have  $ G_\infty(\x,\x';
 \beta)  \leq  \cst'_3 G_\infty (\x,\x';
 16\beta) $.

Set $\y_0:=\x$. Then \eqref{diam}, \eqref{thm4130} together with \eqref{thm412} and
\eqref{thm413}  give
\begin{eqnarray} \nonumber
 \vert F_{j, \infty}^m-F_{j,L}^m \vert \leq  \cst_7
 {\cst'}_3^{j-1}\frac{ \vert \Fl_j \vert^m}{m!}G_\infty(\y_0,\y_1;32(\beta-\tau_1))
... \\
   G_\infty(\y_j,\y_0;32\tau_j)\sum_{l=0}^{j} (2\chi(\x_l) 
  + e^ {-\frac{d^2(\x_l)}{64\beta}}).
\end{eqnarray}
Thus from  this inequality and \eqref{estFL}, we need to estimate the quantity 
\begin {equation} \nonumber 
 Q:= (\sum_{l=1}^{j-1} \sum_{l'=1}^{l} \vert \y_{l'-1}-\y_{l'}\vert \vert \y_{l}-\y_{l+1} \vert)^m
G_\infty(\y_0,\y_1;32(\beta-\tau_1))...G_\infty(\y_j,\y_0;32\tau_j).
\end{equation}
By using \eqref{kahas2} we have 
\begin {equation} \nonumber 
  Q \leq (8 \beta j^2)^{m} m^m 2^{3(j+1)/2}
G_\infty(\y_0,\y_1;64(\beta-\tau_1)).......G_\infty(\y_j,\y_0;64\tau_j).
\end{equation}
and then  for  $j \leq n$
\begin{eqnarray} \label{thm4131}
 \vert F_{j, \infty}^m-F_{j,L}^m \vert \leq 2^{3(n+1)/2}\cst_7
 {\cst'}_3^{n-1}\frac{(8\beta j^2)^{m}m^m}{m!}G_\infty(\y_0,\y_1;64(\beta-\tau_1))... \\ \nonumber
 G_\infty(\y_j,\y_0;64\tau_j)\sum_{l=0}^{j} (2\chi(\x_l) 
  + e^ {-\frac{d^2(\x_l)}{64\beta}}).
\end{eqnarray}
By extending the integration
with respect to $\y_0,...\y_{l-1},\y_{l+1}.. \y_j$ on the whole $\R^3$ space, 
and  using the semigroup property  \eqref{estnoy} and
 the fact that $G_\infty(\x,\x;t)= \frac{1}{(2\pi t)^{3/2}}$ we get
\begin{align}\label{g2121}
\left \vert \int_{\Lambda_L^{j+1}}d\y (F_{j, \infty}^{n-k}-F_{j,L}^{n-k})\right \vert &\leq  c^{n}\cst_7
 {\cst'}_3^{n-1}(j+1)\frac{(\beta j^2)^{{(n-k)}}(n-k)^{n-k}}{\beta
   ^{3/2}(n-k)!} \\ &\cdot  
 \int_{\Lambda_L}d\x (2\chi(\x)  +  e^ {-\frac{d^2(\x)}{64\beta}})
 \nonumber 
\end{align}
 for some positive constant $c$. Moreover, simple estimates show that 
$$\int_{\Lambda_L}d\x (2\chi(\x)  +  e^ {-\frac{d^2(\x)}{64\beta}}) \leq c L^2( 1  + \sqrt \beta),$$ 
where $c$ is also some positive numerical constant.  From \eqref{def-fLN} and  the Theorem \ref{thm3},
\begin{align}  \label{thm414}
& \int_{\Lambda_L}d\y_0 \left\{ \frac{\partial^n G_L}{\partial\omega^n}(\y_0,\y_0;\beta,\omega)-f_L^n(\y_0,\beta,\omega)
 \right\} \\
&= n!\sum_{k=1}^{n}\sum_{j=1}^{k}(-1)^{j}
\sum_{(i_1,...i_j)\in\{1,2\}^{j}} \chi_{j}^{k}(i_1,...,i_j) \nonumber \\ 
&\cdot \int_{D_j(\beta)}d\tau \int_{\Lambda^{j+1}}d\y (F_{j, \infty}^{n-k}-
F_{j,L}^{n-k})(\y_0,\y_1,...,\y_j,\tau_1,...,\tau_j,\beta,\omega).\nonumber
\end{align}
 Then   \eqref{g2121} together with \eqref{thm414} lead to:
$$
\left \vert \int_{\Lambda_L}d\y_0  [\frac{\partial^n G_L}{\partial\omega^n}(\y_0,\x;\beta,\omega)-f_L^n(\y_0,\beta,\omega)]
 \right \vert \leq   L^2 c(n)\frac{(1+ \beta)^{7n+3}}{\sqrt \beta}(1+ \omega)^{3n+2} $$
where $c(n)= (n+1)!
 c^{n}\sum_{k=1}^{n} \frac {(n-k)^{n-k}}{(n-k)!}\sum_{j=1}^{k}  
\frac{ j^{2(n-k)}}{j!}$ and $c$ is again a numerical factor. 
This last estimate clearly  implies \eqref{thm411}.

Let us now prove that for all   $\beta > 0$ and $\omega\geq 0$,  $g_L^n(\y_0,\beta,\omega)$ given in \eqref{def-gLN} satisfies:
\begin{eqnarray}\label{majoresid}
 \vert \int_{\Lambda_L}d\y_0\, g_L^n(\y_0,\beta,\omega)|\leq\,L^2\,g(n,\beta,\omega),
\end{eqnarray}
 where $g(n,\beta,\omega):= c(n)(1+\beta)^{7n +2}(1+\omega)^{3n+2}$ and $c(n)$ is a 
positive constant depending only on $n$. The same arguments
 as above leading to the estimate \eqref{thm4131} imply 
\begin{align}\label{thm415}
& \vert F_{j, \infty}^m(\y_0,...,\y_j,\tau_1,...,\tau_j,\beta,\omega)
\vert \\
&\leq  \cst_7
 {\cst'}_3^{j-1}\frac{ \vert \Fl_j \vert^m}{m!}G_\infty(\y_0,\y_1;32(\beta-\tau_1))
... G_\infty(\y_j,\y_0;32\tau_j) \nonumber \\
&\leq \cst_7
 {\cst'}_3^{j-1}\frac{(8\beta j^2)^{m}m^m 2^{3(j+1)/2}}{m!}G_\infty(\y_0,\y_1;64(\beta-\tau_1))... 
 G_\infty(\y_j,\y_0;64\tau_j).\nonumber 
\end{align}
On the other hand, by the semigroup property (put
$\Lambda_L^c:=\R^3\backslash \Lambda_L$),
\begin{align}\label{thm416}
&\int_{\R^3}d\y_1
...\int_{\Lambda_L^c}d\y_{l}\int_{\R^{3}}d\y_{l+1}...\int_{\R^{3}}d\y_j\,G_\infty(\y_0,\y_1;64(\beta-\tau_1))... 
 G_\infty(\y_j,\y_0;64\tau_j)\nonumber \\
&= \int_{\Lambda_L^c}d\y_l G_\infty(\y_0,\y_l;64(\beta-\tau_l))
 G_\infty(\y_l,\y_0;64\tau_l).
\end{align}
Then \eqref{def-gLN},  \eqref{thm415} and \eqref{thm416} imply
\begin{align}
&\left \vert \int_{\Lambda_L}d\y_0\, g_L^n(\y_0,\beta,\omega)\right
\vert \leq n!\sum_{k=1}^{n}\sum_{j=1}^{k} 
\cst_7
 {\cst'}_3^{j-1}2^{(5j+3)/2}\frac{(8\beta
   j^2)^{(n-k)}(n-k)^{n-k}}{(n-k)!} \nonumber \\  
&\cdot \int_{D_j(\beta)}d\tau 
  \sum_{l=1}^{j}\int_{\Lambda_L}d\y_0\int_{\Lambda_L^c}d\y_{l} G_\infty(\y_0,\y_l;64(\beta-\tau_l))
 G_\infty(\y_l,\y_0;64\tau_l).
\end{align}
By using the explicit form of the heat kernel given in \eqref{heatk}, a straightforward computation shows that
$$\int_{\Lambda_L}d\y_0 \int_{\Lambda_L^c}d\y_{l} G_\infty(\y_0,\y_l;64(\beta-\tau_l))
 G_\infty(\y_l,\y_0;64\tau_l) \leq c  \frac {L^2} {\beta}$$
 for some positive constant $c$. Hence we get
\begin{eqnarray}\nonumber
\vert \int_{\Lambda}d\y_0\, g_L^n(\y_0,\beta,\omega)|\leq L^2 c(n)(1+\beta)^{7n +2}(1+\omega)^{3n+2}
\end{eqnarray}
 where $c(n)= n!c^n \sum_{k=1}^{n}\sum_{j=1}^{k}
 \frac{ j^{2(n-k)}(n-k)^{n-k}}{(n-k)!(j-1)!}$ and $c$ is a  positive  numerical factor.  This    shows \eqref{majoresid}. 
Then \eqref{majoresid} and \eqref{thm411} imply the theorem. \qed 

\vspace{0.5cm}

\subsection { The proof of Theorem \ref{infinitevolume}}

We are now ready to prove the thermodynamic limit of generalized
susceptibilities in the grandcanonical ensemble, when the chemical
potential is negative (fugacity $z$ less than one). 

Let $L \geq 1$, $\beta >0$, $\omega \geq 0$ and $\vert z\vert <1$. 
We know from \eqref{Pinfini} and \eqref{PL}  that:

\begin{align} 
& P_L (\beta,\omega,z,\epsilon) -  P_\infty (\beta,\omega,z,\epsilon)
\\ &=  \frac{\epsilon}{\beta \vert  \Lambda_L\vert
}\sum_{k\geq 1} \frac{(-\epsilon z)^k }{k} \int_ {\Lambda_L} d\x   \{
G_L(\x,\x;k\beta,\omega) -  G_\infty(\x,\x;k\beta,\omega) \}.
\nonumber
\end{align}
Then by applying the Theorem \ref{thm41} we get 
$$  \frac{\partial^n (P_L-P_\infty )}{ \partial
\omega^n} = \frac{\epsilon}{\beta \vert  \Lambda_L\vert
}\sum_{k\geq 1} \frac{(-\epsilon z)^k }{k}  \int_ {\Lambda_L} d\x \left( \frac{\partial^n G_L }{ \partial
\omega^n} - \frac{\partial^n G_\infty}{ \partial \omega^n} \right)(\x,\x;k\beta,\omega).
 $$
In particular, this also shows that the series from \eqref{chiinfini}
must converge.  
 Moreover  by  using   again  the bound \eqref{majodiff}
in the last formula, we
have

$$ \vert\chi_L^{(n)} - \chi_\infty^{(n)}  \vert \leq c(n) (1+
\omega)^{3n +5}\frac{1}{ \beta L}\sum_{k\geq 1}
\frac {\vert z \vert^k} {k} \frac{(1+ k\beta)^{7n+8}}{ \sqrt{k\beta}}.  $$
Since the series in the r.h.s of this last inequality is finite and
$L$ independent,  this  proves \eqref{limite}. \qed 

\vspace{0.5cm} 

\noindent {\bf Acknowledgments.} The authors thank V. A. Zagrebnov, G.
Nenciu and N. Angelescu for many fruitful discussions. H.C. was
partially supported by the embedding grant from {\it The Danish
  National Research Foundation: Network in Mathematical Physics and
  Stochastics}. H.C. also acknowledges support from the Danish 
F.N.U. grant {\it  Mathematical Physics and Partial Differential
  Equations}, and partial support through the European Union's IHP
network Analysis $\&$ Quantum HPRN-CT-2002-00277.


\end{document}